# A Specialized Importance-Aware Quantum Convolutional Neural Network with Ring-Topology (IA-QCNN) for MGMT Promoter Methylation Prediction in Glioblastoma

Emine Akpinar[1*], Murat Oduncuoglu[1]

[1]Department of Physics, Yildiz Technical University, Istanbul, Turkey

*Corresponding author: Emine Akpinar, emineakpinar28@gmail.com

## Abstract

Glioblastoma (GBM) is a highly aggressive primary malignancy in adults, necessitating personalized therapeutic strategies due to its inherent molecular heterogeneity. MGMT promoter methylation serves as a pivotal prognostic and predictive biomarker for anticipating clinical response to temozolomide-based chemotherapy. Although various AI frameworks have been developed for non-invasive MGMT prediction, spatial heterogeneity of methylation status and the high-dimensional, pixel-intensive, and highly correlated nature of MRI data frequently constrain discriminative feature learning and generalizability of classical architectures. To circumvent these limitations, a specialized Importance-Aware Quantum Convolutional Neural Network with Ring-Topology (IA-QCNN) architecture is proposed, based on the principles of quantum mechanics, including superposition and entanglement, and enabling more efficient representation learning in high-dimensional Hilbert space. The framework establishes a methodological bridge between GBM radiogenomics and quantum deep learning by integrating energy-based slice selection, importance-aware weighting, ring-topology quantum convolution, and folding-based pooling layers. When the model predicts MGMT promoter methylation status using both mpMRI and T1Gd images, experimental results demonstrate that the IA-QCNN achieves high accuracy despite its low number of trainable parameters while effectively minimizing the overfitting problem observed in classical models. Quantitative analyses reveal that the T1Gd modality possesses higher discriminative power than mpMRI, establishing a clinically significant sequence preference. Furthermore, the model exhibits exceptional robustness in hybrid noise environments, effectively utilizing noise as a regularization mechanism to enhance predictive performance. Consequently, the specialized IA-QCNN architecture provides a robust, reliable, and computationally efficient alternative to classical approaches in the analysis of heterogeneous radiogenomic data, establishing a concrete foundation for quantum advantage.

## 1. Introduction

Malignancies of the central nervous system constitute one of the most challenging groups of diseases in oncology owing to their intrinsic biological complexity. Among this group, glioblastoma (GBM), also referred to in the literature as glioblastoma multiforme, is considered the most aggressive malignant brain tumor with the poorest prognosis among gliomas originating from glial cells [1–5]. Among all malignant brain tumors, GBM is the most frequently observed, accounting for 50.1% of cases [6], with a reported annual incidence of approximately 3.19 per 100,000 individuals [7]; this rate is thought to be increasing in association with population aging and improvements in diagnostic methods [4,8]. Although GBM can occur at any age, it is more frequently observed in the adult population [9–11] and demonstrates a higher incidence in males compared to females [10,11]. In terms of anatomical localization, GBM predominantly affects the cerebral hemispheres in adults and is classified into two main groups, primary and secondary, based on its clinical and molecular characteristics [12].

Primary (*de novo*) GBMs arise via multistep tumorigenesis from normal glial cells and account for approximately 90% of all GBM cases, whereas secondary GBMs develop through the molecular progression and malignant transformation of low-grade gliomas, exhibiting slower growth and a more favorable prognosis compared to their primary counterparts [9,11–14]. In 2016, in the World Health Organization (WHO) classification of the central nervous system, molecular alterations were incorporated into the diagnostic evaluation process of certain tumors in addition to histopathological findings, and the published report emphasized that GBMs exhibit a molecularly heterogeneous structure [15]. In 2021, the WHO CNS classification placed molecular features at the center of the final integrated diagnostic process; thus, the decisive role of molecular genetics in the diagnosis of GBM was clearly established [16]. Accordingly, GBM (WHO grade IV) is the most aggressive diffuse glioma of astrocytic origin; in adults, IDH-wildtype diffuse astrocytic gliomas are defined by specific histopathological and molecular criteria [17]. Advances in the molecular classification of GBMs have increased interest in more complex and multimodal treatment approaches and have revealed a clear shift toward precision medicine. Currently, in the treatment of GBM, a multimodal approach is applied that includes surgical resection followed by radiotherapy and concomitant chemoradiotherapy with temozolomide, taking into account factors such as the patient's age, time of diagnosis, tumor localization, and performance status. In this context, according to the most recent implementation of the WHO classification, $O^6$-methylguanine-DNA methyltransferase (MGMT) promoter methylation serves as a robust predictive biomarker for estimating treatment response and temozolomide (TMZ)-associated survival outcomes in GBM; it is also recognized as a key indicator in the evaluation of treatment strategies involving TMZ administered either alone or in combination with radiotherapy [1,18–20].

The MGMT gene encodes an enzyme involved in DNA repair, establishing a protective mechanism capable of reversing the cytotoxic and DNA-damaging effects of alkylating and methylating chemotherapeutic agents on tumor cells [21–23]. These agents specifically target the $O^6$ position of guanine bases in tumor cell DNA, leading to the addition of alkyl groups and consequent DNA damage [1]. In cases where the MGMT promoter is unmethylated (corresponding to active MGMT), ongoing MGMT-mediated DNA repair contributes to the development of resistance to temozolomide and limits clinical benefit; conversely, MGMT promoter methylation suppresses gene expression, functionally inactivates MGMT, reduces DNA repair capacity, and enhances the efficacy of temozolomide treatment, thereby correlating with improved prognosis [24–27]. In contemporary clinical studies, MGMT promoter methylation status is utilized as a fundamental biomarker for predicting therapeutic response, playing a particularly important role in determining first-line treatment strategies for elderly GBM patients [25]. In other words, MGMT status serves as a primary guiding factor in selecting the most appropriate treatment modality for elderly patients, whether it involves radiotherapy alone, temozolomide chemotherapy, or combined TMZ/radiotherapy.

Moreover, recent studies in the literature have reported that an unmethylated MGMT promoter is associated with increased tumor blood volume [1], highlighting the potential role of magnetic resonance imaging (MRI)-based imaging biomarkers in predicting therapeutic response and prognosis [28]. In addition, proton MR spectroscopy can detect 2-hydroxyglutarate (2-HG) levels, which are associated with IDH1/2 mutations [29]; other MRI-derived parameters utilized to infer genetic alterations include apparent diffusion coefficient (ADC) values, the ratio of T2-contrast enhancing volume, and relative cerebral blood volume. Furthermore, T1-weighted contrast-enhanced (T1-CE) MRI emerges as a preferred non-invasive imaging biomarker for GBM, widely employed not only in prognostic assessment but also in diagnostic and therapeutic workflows [4]. GBMs are characterized by ill-defined borders, infiltrative growth, and pronounced heterogeneity, with blood–brain barrier disruption closely linked to contrast agent (gadolinium) uptake [30]. Consequently, in T1-CE MRI, contrast enhancement becomes pronounced in regions of the brain where the blood–brain barrier is compromised due to tumor infiltration. Currently, in the field of radiogenomics, multiple methodologies and models are being developed to accurately characterize the genetic features of tumors and guide appropriate treatment strategies using imaging data derived from multiple MRI sequences, such as T1-CE, with the aim of reducing the need for surgical intervention and shortening the time required for genetic characterization [28,31]. Accordingly, there is growing interest in non-invasive alternatives to the high-cost and invasive molecular tests traditionally used to determine MGMT promoter methylation status [32]. However, the high degree of vascularization, infiltrative growth, aggressive clinical course, and pronounced genetic and phenotypic heterogeneity of GBMs complicate the analysis of their high-dimensional and complex imaging data (e.g., radiomics features), necessitating advanced computational and artificial intelligence (AI)-based approaches [33]. In particular, the short survival times observed in elderly individuals underscore the need for rapid and accurate diagnosis and prognostication, enabling the implementation of personalized and targeted treatment strategies. Therefore, there is a critical need for computer-aided diagnosis (CAD) systems that provide non-invasive prediction of MGMT promoter methylation status from MRI images and can serve as a second-opinion tool. Notably, convolutional neural network (CNN)-based deep learning (DL) architectures are widely employed in such predictive tasks due to their capacity for automatic feature extraction and high classification accuracy in medical imaging applications.

Currently, CNN-based deep learning (DL) models with diverse architectures are being actively investigated for tasks such as disease detection, segmentation, and classification [34–36], as well as for their potential contributions to personalized treatment planning [37], aiming to support therapeutic decision-making and improve patient outcomes through non-invasive approaches [38]. Compared to classical machine learning (ML) models that fundamentally rely on handcrafted features, DL architectures offer distinct advantages, such as the ability to automatically extract features from imaging data like MRI and the capacity to efficiently process large-scale datasets. CNNs are DL models developed based on the structure of the visual cortex in the human brain, utilizing a hierarchical learning approach in which each feature level is constructed upon lower-level features extracted by the previous layer, allowing the model to learn complex structures and relationships from simple concepts [39–41]. A typical CNN model consists of input, convolution, activation, pooling, and fully connected layers, wherein feature learning is generally performed by small filters at each layer designed to detect specific patterns. Moreover, CNN architectures exhibit a certain degree of robustness and invariance to spatial translations, rotations, and noise, owing to their structural characteristics [42]. In the literature, numerous CNN-based DL architectures have been developed to predict MGMT promoter methylation status, a molecular feature that significantly influences treatment response and clinical outcomes in GBM patients, directly from MRI images. In these studies, CNN-based models are often trained using conventional MRI sequences such as T1, T1-CE, T2, and FLAIR, as well as multiparametric MRI (mpMRI) data; in some studies, advanced MRI sequences including diffusion-weighted imaging (DWI), apparent diffusion coefficient (ADC), and arterial spin labeling (ASL) are also incorporated into the models [43–47]. Additionally, radiomics-based approaches leveraging high-dimensional imaging features have been employed in certain cases [48]. However, there are two fundamental challenges encountered in predicting MGMT promoter methylation status from MRI data using CNN-based DL models:

- The spatially heterogeneous and variable nature of MGMT status within GBM tumor structures complicates the prediction process. In this context, the accurate representation of input data is critical for adequately capturing the biological diversity of the tumor.
- The high-dimensional, pixel-intensive, and highly correlated structure of MRI data can limit the ability of classical CNN models to discriminate subtle textural differences and to learn optimal feature representations. Increasing model complexity to enhance discriminative power, however, leads to higher computational costs, increased energy consumption, and potential generalization issues.

These limitations have incentivized the development of more efficient and representationally powerful alternatives beyond classical CNN-based DL models. In this regard, Quantum Convolutional Neural Networks (QCNNs), which integrate principles of quantum computing with convolutional learning structures, have emerged as particularly promising approaches.

Upon examination of the existing literature, no study has been identified in which QCNN-based approaches have been directly applied to predict MGMT promoter methylation status from MR images. Nevertheless, various studies have reported that QCNN and, more broadly, quantum machine learning (QML) / quantum neural network (QNN) architectures offer promising alternatives for problems requiring the analysis of complex biomedical data [49–53]. In particular, successful applications of quantum and hybrid quantum–classical learning approaches have been documented in tasks such as brain tumor classification, neuroimaging-based diagnosis, and Alzheimer's disease staging [54–59]. Fundamentally, QCNN is a quantum circuit architecture developed by drawing inspiration from classical convolutional neural networks. While QCNN aims to construct data representations by learning low-level local features from input data, similar to classical CNN architectures, it simultaneously enables richer modeling of complex data structures by leveraging fundamental quantum principles such as superposition and entanglement, thereby offering potential advantages in advanced feature extraction [60–62]. Through the property of superposition, in which the amplitudes of all qubit states are manipulated simultaneously, multiple probabilistic states can be represented concurrently; the property of entanglement, on the other hand, enables the joint modeling of correlation structures among variables within the circuit [63,64]. In addition, the Hilbert space, in which quantum states are defined, provides a higher representational capacity and a richer feature space compared to classical vector spaces [65,66]. Within this framework, QCNN preserves the fundamental characteristics of classical CNNs, such as measurement-induced nonlinearity, local convolutional structure, and scalability to deep architectures; at the same time, owing to architectural properties such as shallow circuit depth requirements, relative noise tolerance, and the mitigation of input–output bottlenecks, it is considered among the most feasible approaches for implementation on noisy intermediate-scale quantum (NISQ) devices. At present, multiple QCNN architectures have been developed based on different methodological approaches. Among the QCNN architectures proposed in the literature, the model developed by Cong et al., which adapts the convolution–pooling structure of classical CNNs to quantum circuits and enables logarithmic parameter scaling ($O(\log(N))$) along with NISQ-compatible training [67], and the quanvolutional approaches proposed by Henderson et al., which apply randomly parameterized quantum circuits to local data subregions, are particularly notable [64]. Furthermore, hybrid quantum–classical QCNN (QCCNN) architectures, in which classical convolution operations are implemented using parameterized quantum circuits, have been reported to yield promising results in terms of circuit depth analysis and learning stability under noise [68]. In addition, in QCNN architectures frequently employed in classification problems in the literature, the "shared weight mechanism" inherent to classical CNN structures is generally defined based on qubit positional information (particularly low-significance position qubits) [69]. However, this approach may not always yield optimal results for highly correlated, structurally complex, and heterogeneous neuroimaging data (e.g., MRI scans); representations of qubits discarded as insignificant or assigned low weights may implicitly contain clinically critical information, such as tumors, lesions, or subtle structural abnormalities. Therefore, allowing the model to learn the weighting process during training both reduces user-induced bias and ensures that clinically significant local patterns are automatically emphasized.

In this study, an" importance-aware weighting" approach is proposed that goes beyond the "position-based weight-sharing" mechanisms commonly employed in the literature, by scaling the contribution of features derived from pixel intensity information to their quantum representation through learnable parameters. Consequently, instead of position-based weighting, a data-driven weighting strategy dependent on feature values is adopted. In other words, the conventional concept of attention-based weighting is adapted to an angle-based quantum feature encoding framework without performing direct amplitude encoding. The proposed importance-aware weighting approach is integrated into a *specialized Importance-Aware Quantum Convolutional Neural Network with Ring-Topology (IA-QCNN)* architecture at the stage of quantum feature encoding, thereby presenting a tailored and end-to-end learnable model for predicting MGMT promoter methylation status from MRI images in GBM tumors.

The contribution of this study to the literature can be summarized as follows:

- To the best of our knowledge, this study presents the first comprehensive quantum-based approach specifically designed to predict MGMT promoter methylation status directly from MRI images in GBM tumors. Through the problem-specific development of a specialized Importance-Aware Quantum Convolutional Neural Network with Ring-Topology (IA-QCNN) architecture, a direct methodological bridge is established between GBM radiogenomic analysis and quantum deep learning (DL) techniques.

- Furthermore, this study proposes an "importance-aware weighting" strategy, as an alternative to the position-based weighting approaches commonly employed in the literature. This strategy scales the contribution of features derived from pixel intensity information to their quantum representations through learnable and adaptive parameters at the quantum feature encoding stage. By integrating the importance-aware weighting strategy into the proposed specialized IA-QCNN architecture, a data-driven, end-to-end learnable model is realized, which can more effectively highlight discriminative local patterns.

- The performance of the proposed specialized IA-QCNN architecture in predicting MGMT promoter methylation status from MRI images of GBM tumors was evaluated using the RSNA-MICCAI Brain Tumor Radiogenomic dataset. Model performance was analyzed through fundamental classification metrics—including precision, recall, and F1-score—alongside mean training/validation loss and accuracy. The findings demonstrate that the model possesses a high learning capacity.

- Comparative experiments conducted across multi-parametric MRI sequences (FLAIR, T1w, T1Gd, and T2) revealed that T1-weighted contrast-enhanced (T1Gd) images yielded superior model performance in predicting MGMT promoter methylation. This finding suggests that intensity variations associated with contrast uptake are more effectively represented by the angle-based quantum feature encoding, providing supportive evidence for a clinically significant sequence selection.

- The performance of the proposed specialized IA-QCNN architecture was further evaluated under mild-to-moderate Gaussian noise across three scenarios: (i) noise added solely to the image data, (ii) noise applied only to quantum gate operations, and (iii) hybrid noise applied simultaneously to both image data and quantum gates. In the first two scenarios, only a moderate decrease in classification performance was observed, indicating that the model exhibits relative robustness to noise. Notably, under hybrid noise conditions, the model appeared to exploit noise as a form of regularization mechanisim, resulting in performance enhancement.

- Finally, when compared to CNN/DNN models with similar parameter counts, transfer learning (TL) approaches, and state-of-the-art deep learning (SOTA DL) models, the proposed specialized IA-QCNN architecture achieved higher predictive accuracy with significantly fewer parameters. Moreover, when benchmarked against studies in the literature reporting MGMT promoter methylation prediction, the model demonstrated competitive performance and offers a computationally efficient alternative to classical approaches.

The remainder of this study is organized as follows:

**Section 2** is divided into three primary subsections. The first subsection introduces the RSNA-MICCAI Brain Tumor Radiogenomic dataset utilized in this work and provides detailed technical information regarding the mpMRI sequences. The second subsection describes the preprocessing steps applied to the MRI images, including the proposed Energy-Based Slice Selection method. The final subsection of **Section 2** elaborates on the mathematical foundations and physical principles underlying the proposed specialized Importance-Aware Quantum Convolutional Neural Network with Ring-Topology (IA-QCNN) architecture. The section concludes by presenting the evaluation metrics employed to assess the classification performance of both the proposed model and the classical DL models used for comparative purposes. **Section 3** examines the findings obtained using the proposed specialized IA-QCNN architecture for predicting MGMT promoter methylation status in GBM tumors, organized into six subsections. The first two subsections present the results obtained from T1Gd and mpMRI images, respectively, while the subsequent three subsections analyze the performance of the proposed model under three distinct noise scenarios. The final subsection provides a comparative analysis of CNN/DNN, TL, and SOTA models executed under identical experimental conditions. **Section 4** offers a comprehensive discussion and analysis of the numerical results presented in **Section 3** in the context of the existing literature. Finally, **Section 5** summarizes the main findings of the study, evaluates the limitations of the proposed method, and outlines directions for future research.

2. **Materials and Method**

In this section, a specialized Importance-Aware Quantum Convolutional Neural Network with Ring-Topology (IA-QCNN) architecture for predicting the MGMT promoter methylation status of GBM tumors from MRI images is described in detail. The overall workflow of the model is presented in **Figure 1**. The first part introduces the RSNA-MICCAI Brain Tumor Radiogenomic dataset and provides detailed information on the MRI sequences employed. The second part outlines the preprocessing steps applied prior to the training and testing phases; subsequently, the proposed IA-QCNN architecture is elaborated in depth, with particular emphasis on its quantum and classical computational components.

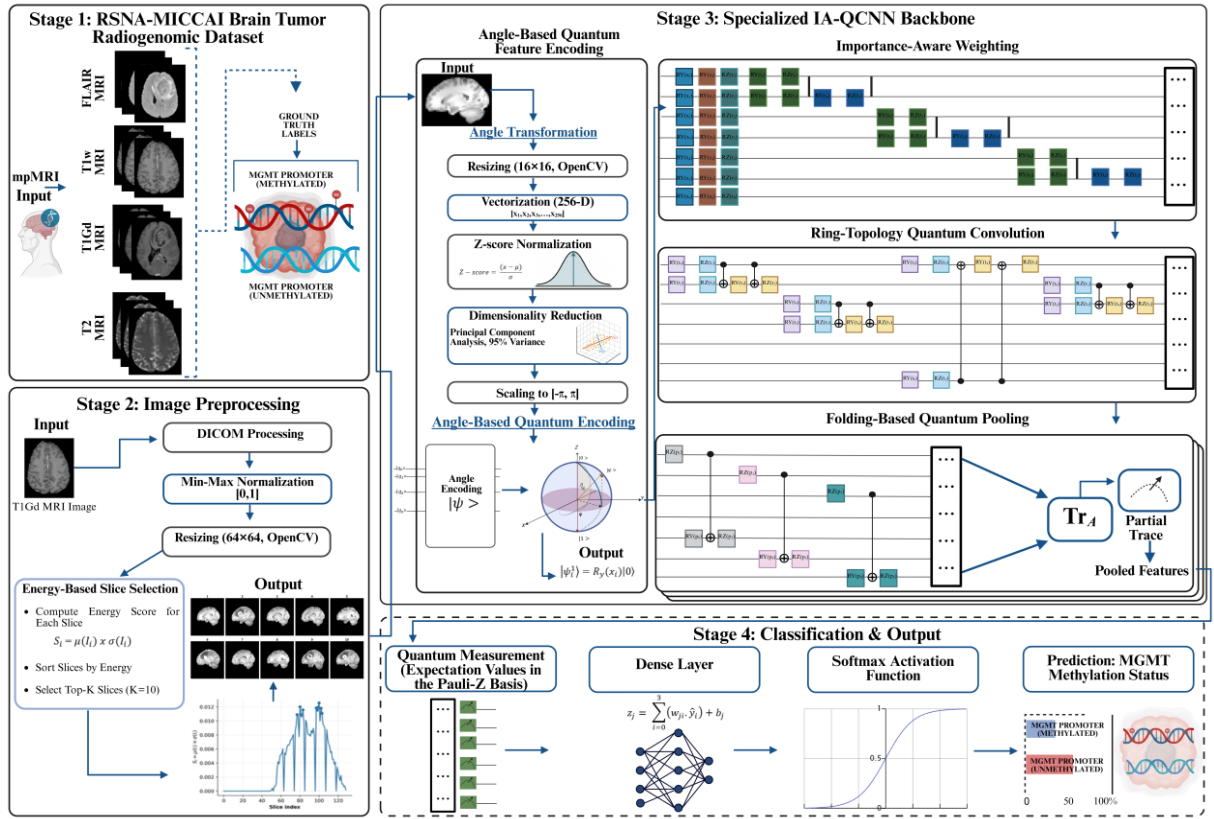

**Figure 1.** General workflow of the proposed study. In the first stage, the RSNA-MICCAI Brain Tumor Radiogenomic dataset and the mpMRI sequences employed in this study are introduced. The second stage outlines the DICOM preprocessing, normalization, and rescaling steps applied to the mpMRI images, with particular emphasis on T1Gd MRI, alongside the proposed "Energy-Based Slice Selection" method. In the third stage, a specialized Importance-Aware Quantum Convolutional Neural Network with Ring-Topology (IA-QCNN) architecture for predicting the MGMT promoter methylation status of GBM tumors directly from MRI images is presented. In this section, angle-based quantum feature encoding, importance-aware weighting, ring-topology quantum convolution, and folding-based quantum pooling layers are described in detail from both physical and mathematical perspectives. In the final stage, measurement outcomes in the Pauli-Z basis are passed to a dense layer followed by a softmax activation function, yielding class probabilities and final prediction outputs.

### 2.1. RSNA-MICCAI Brain Tumor Radiogenomic Dataset and Multiparametric (mp) MRI Sequences

In 2021, the Radiological Society of North America (RSNA) and the Medical Image Computing and Computer Assisted Interventions Society (MICCAI) collaborated to determine the MGMT promoter methylation status in GBM tumors from MRI images, with the aim of improving diagnosis and treatment planning for GBM patients. This collaboration led to a global challenge entitled Brain Tumor Radiogenomic Classification. The dataset provided for the challenge has been made publicly available, comprising multiparametric (mp) MRI scans for each GBM patient. All images underwent skull-stripping, were de-identified, and stored in DICOM format. For each patient, four imaging modalities were acquired: FLAIR (Fluid Attenuated Inversion Recovery), pre-contrast T1-weighted (T1w), post-contrast T1-weighted (T1Gd), and T2-weighted (T2). The mpMRI images for each patient were organized in separate folders, each labeled with a unique five-digit identifier. Each modality provides distinct information regarding the tumor and surrounding brain tissue and consists of MRI sequences obtained within a defined time interval. Furthermore, the dataset categorizes patients according to MGMT status into two classes: MGMT-methylated (1) and MGMT-unmethylated (0) [70].

The characteristics of the imaging modalities employed are detailed as follows [71,72]:

**FLAIR MRI:** A specialized T2-weighted sequence that suppresses cerebrospinal fluid signals, enabling clearer visualization of lesions and certain pathological alterations. It is particularly useful for detecting lesions adjacent to the ventricles, identifying small lesions, delineating edema and infiltrative regions, and revealing abnormalities that may not be evident in other modalities.

**T1w MRI:** Highlights the anatomical details of the brain and generates distinctive contrast between tissues with varying water–fat compositions. Tissues with high water content appear darker, whereas fat-rich tissues appear brighter.

**T1Gd MRI:** Following the administration of an intravenous gadolinium-based contrast agent, certain regions become more conspicuous on MRI images. Post-contrast T1Gd MRI highlights tumor regions with disrupted blood-brain barriers as bright, contrast-enhancing areas. These hyperintense regions typically correspond to biologically active, highly vascularized portions of the tumor. In other words, T1Gd MRI plays a critical role in visualizing the biologically active regions of the tumor. As emphasized in numerous clinical studies, contrast-enhanced MRI remains one of the most frequently employed imaging modalities for GBM diagnosis.

**T2 MRI:** Provides crucial information for delineating tumor boundaries and surrounding cerebral edema. Increased water content prolongs T2 relaxation times, thereby assisting in differentiating the tumor mass from adjacent tissue.

The radiogenomic classification dataset used in this study consists of mpMRI scans from 585 GBM patients. The dataset exhibits a balanced class distribution, comprising 307 cases in the MGMT-methylated (1) class and 278 cases in the MGMT-unmethylated (0) class. The MGMT class distribution, the number of images per MRI modality, and the orientation-based distribution within each modality (axial, coronal, and sagittal) are presented in detail in **Figure 2a–c**.

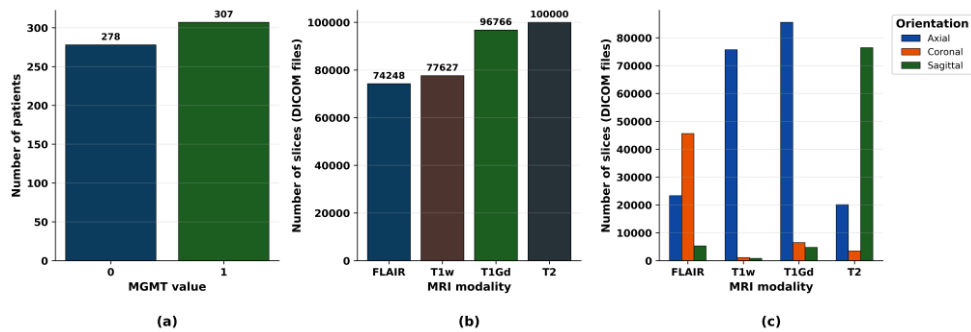

**Figure 2.** Distribution of the RSNA–MICCAI Brain Tumor Radiogenomic dataset according to various characteristics. **(a)** Class distribution based on MGMT status, comprising 307 cases with positive (methylated) MGMT promoter status and 278 cases with negative (unmethylated) status. **(b)** Total number of slices calculated across all DICOM files for each MRI modality, with 74,248, 77,627, 96,766, and 100,000 slices for FLAIR, T1w, T1Gd, and T2, respectively. **(c)** Detailed distribution of slices according to axial, coronal, and sagittal orientations for each MRI modality.

Representative images of FLAIR, T1w, T1Gd, and T2 MRI modalities for two cases with positive and negative MGMT promoter methylation status are illustrated in **Figure 3**. Additionally, a sequence of images consisting of T1Gd MRI slices from a representative case is presented in **Figure 4**.

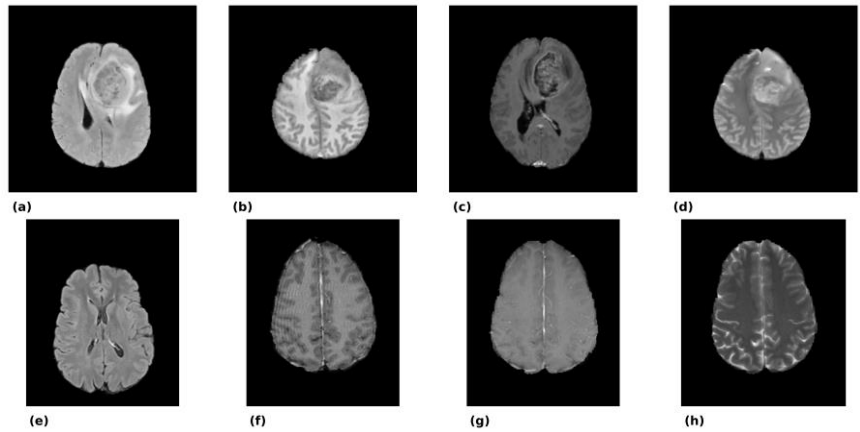

**Figure 3.** Examples of mpMRI modalities from two randomly selected cases with positive and negative MGMT promoter methylation status. Panels **(a–d)** depict FLAIR, T1w, T1Gd, and T2 images from an MGMT-methylated (positive) case, respectively, while panels **(e–h)** illustrate FLAIR, T1w, T1Gd, and T2 images from an MGMT-unmethylated (negative) case.

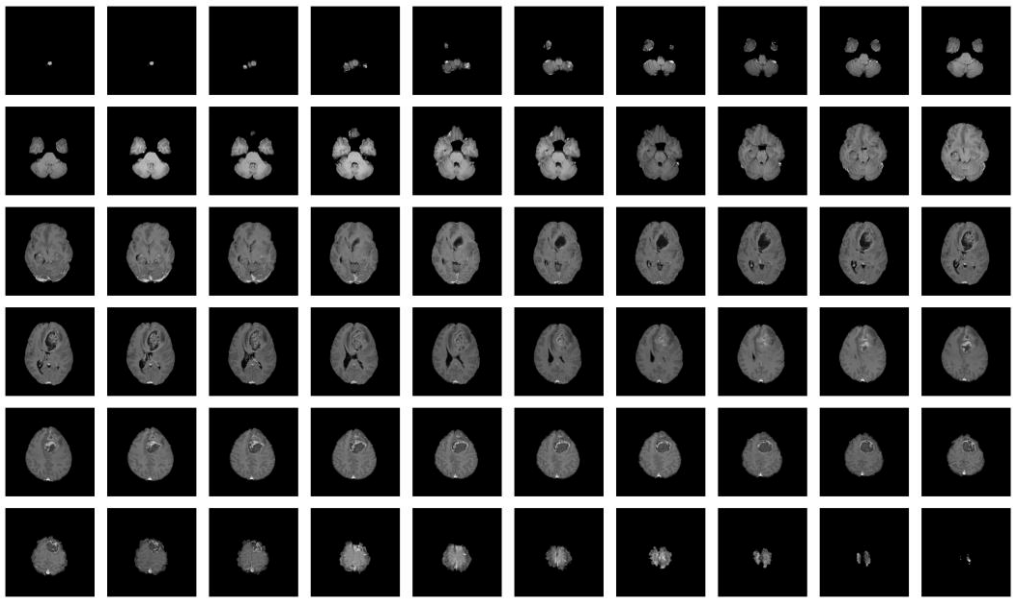

**Figure 4.** Sequential representation of T1Gd MRI slices obtained from Case 146 in the dataset. From the series of 76 slices, images with low information content were filtered out, and only slices containing meaningful anatomical structures were selected for visualization.

### 2.2. Image Preprocessing

Following the dataset definition, all DICOM images corresponding to the mpMRI modalities were subjected to normalization to scale pixel intensity values to a common range. Subsequently, the proposed Energy-Based Slice Selection method was applied.

Detailed descriptions of these procedures are provided below.

#### 2.2.1. DICOM Normalization

As each patient's DICOM files contain pixel data, metadata, and acquisition information, the pixel data were first converted into floating-point matrices. To mitigate intensity variability arising from differences in scanner hardware and acquisition parameters, slice-wise min–max normalization was applied across all DICOM images, scaling pixel intensities to the [0,1] range:

$$x_{normalize} = \frac{(x - \min(x))}{(\max(x) - \min(x))} \tag{1}$$

where $x_{normalize}$ denotes the normalized value, $x$ the original pixel intensity, and $\min(x)$ and $\max(x)$ the minimum and maximum intensities of the corresponding slice, respectively.

Before proceeding to the Energy-Based Slice Selection stage, the normalized images were resized to a fixed target dimension to facilitate energy score computation. Given its suitability for downsampling, the INTER_AREA interpolation method from the OpenCV library was employed.

### 2.2.2. Energy-Based Slice Selection

In this study, the training and testing of the proposed specialized Importance-Aware Quantum Convolutional Neural Network with Ring-Topology (IA-QCNN) architecture were conducted using T1Gd images from the mpMRI modalities as the reference, and the MGMT promoter methylation status was predicted based on this modality. T1Gd MRI, acquired after the administration of contrast agent, is particularly effective in visualizing the active tumor regions and guiding diagnostic and therapeutic decisions, as tumor tissue with a disrupted blood-brain barrier appears hyperintense, or enhanced, on these images [4]. The regions highlighted in T1Gd MRI correspond largely to biologically active, vascularized portions of the tumor, making them clinically relevant reference points [73]. Once the T1Gd MRI images for each patient were identified, slice selection from the DICOM volumes was performed according to the Energy-Based Slice Selection method proposed in this study.

Previous studies in the literature have addressed the utilization of MRI slices from volumetric data without requiring segmentation or manual annotation through various strategies. In some approaches, a fixed number of slices is selected from the central region of the volume as model input, whereas in others, each slice is processed independently, with patient-level predictions derived via majority voting of the slice-level predictions. Furthermore, DL-based approaches have been proposed that treat the MRI volume as an ordered sequence of slices, processing all slices sequentially [32,47,74]. However, such heuristic and pre-defined strategies may limit model reproducibility in complex radiogenomic tasks, such as predicting MGMT promoter methylation status from MR images of GBM tumors, and can increase the risk of anatomical bias.

To address these limitations, an energy-based slice selection strategy is proposed in this study to identify tumor regions rich in information and to select slices with distinctive contrast without requiring segmentation. Initially, following DICOM normalization, the T1Gd MRI images for all patients were resized to a fixed dimension of (64, 64) to reduce processing time and improve computational efficiency. An energy score was then computed for each T1Gd slice, and the slices were ordered such that those with the lowest scores were positioned at the bottom.

The energy score for each slice is defined as:

$$S_i = \mu(I_i) \; x \; \sigma(I_i) \tag{2}$$

where $S_i$ denotes the energy score of the $i^{th}$ slice, $\mu(I_i)$ the mean pixel intensity, and $\sigma(I_i)$ the standard deviation of the corresponding slice. In other words, the energy score of each slice is calculated as the product of its mean intensity and standard deviation, and slices are ranked according to these values. In the subsequent step, the top K slices with the highest energy, representing tumor regions with high contrast and rich information content, were selected based on this ranking. Experimental analyses conducted in this study determined K = 10, and the energy-based slice selection procedure was applied using this value for all patients. This approach enables systematic selection of slices with high signal energy, facilitating the identification of information-rich slices that better represent the tumor region and are more likely to contain radiogenomic information associated with MGMT status. Additionally, this strategy enhances the reproducibility of the model.

**Figure 5** illustrates the workflow of the proposed energy-based slice selection method for a representative GBM patient, while **Figure 6** depicts the selected T1Gd MRI slices highlighting high-contrast, information-rich tumor regions when K = 10. Further example results demonstrating the operation of the energy-based slice selection method for different patients are provided in **Supplementary Figures S1 and S2**.

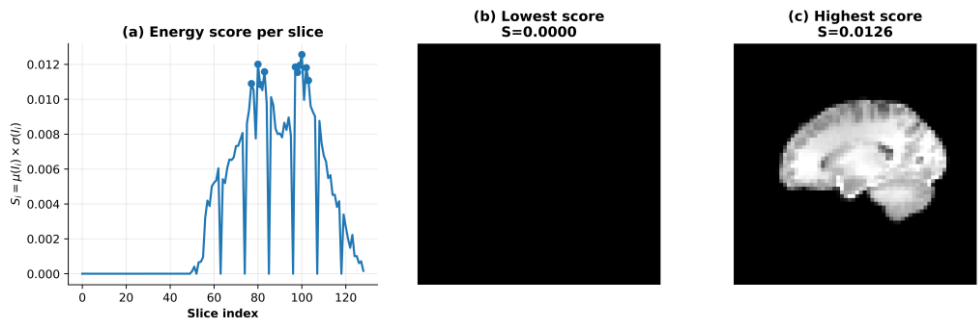

**Figure 5.** Visualization of the proposed energy-based slice selection strategy for Case 00070 from the training dataset. **(a)** Distribution of energy scores for each slice, computed using the approach $S_i = \mu(I_i) \times \sigma(I_i)$, as a function of the slice index. Based on this distribution, the top K = 10 slices with the highest energy scores correspond to slices with indices 100, 80, 99, 97, 102, 83, 98, 103, 77, and 81. **(b)** The slice with the lowest energy score represents regions with minimal anatomical content, predominantly consisting of background information. **(c)** The slice with the highest energy score illustrates the slice selected by the proposed method, in which anatomical structures and clinically informative brain regions are prominently observable.

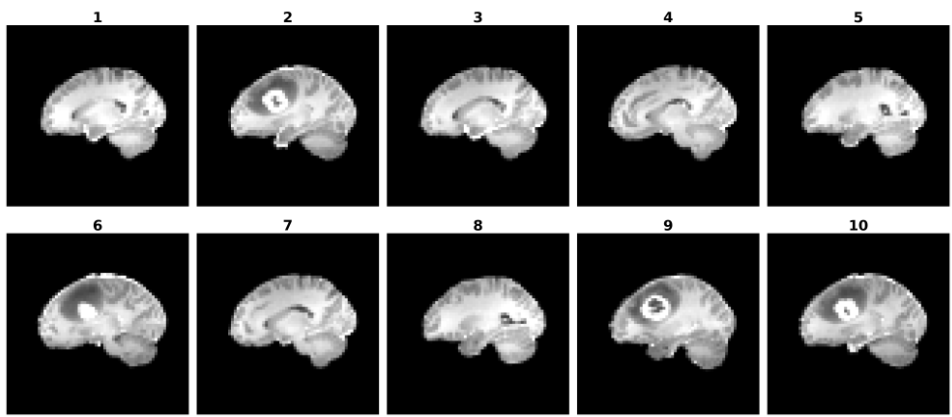

**Figure 6.** The top K = 10 slices selected based on slice-wise energy scores computed using the proposed energy-based slice selection method, along with their corresponding T1Gd MRI representations.

### 2.3. Specialized Importance-Aware Quantum Convolutional Neural Network with Ring-Topology (IA-QCNN) Architecture

In this section, the specialized Importance-Aware Quantum Convolutional Neural Network with Ring-Topology (IA-QCNN) architecture, specifically designed for predicting the MGMT promoter methylation status of GBM tumors directly from MRI images, is described in detail. The proposed IA-QCNN architecture presents a hybrid framework that integrates both quantum and classical computational approaches. The quantum component of the model comprises a QCNN structure, including angle-based quantum feature encoding with an integrated importance-aware weighting strategy, ring-topology quantum convolution, folding-based quantum pooling layers, and measurement operations. The classical component encompasses a dense layer, softmax activation, loss calculation, and optimization procedures. This hybrid architecture is designed to enable end-to-end learning of features extracted from MRI data obtained via the energy-based slice selection method, effectively bridging the quantum and classical components.

### 2.3.1 Angle-Based Quantum Feature Encoding and Importance-Aware Weighting

In this subsection, the angle-based quantum feature encoding and importance-aware weighting mechanisms, which constitute the quantum component of the IA-QCNN architecture, are discussed in detail. This section encompasses angle transformation, angle-based quantum encoding, and importance-aware weighting operations. The algorithmic structure of the angle-based quantum feature encoding and importance-aware weighting approach is illustrated in **Figure 7**.

```
Algorithm 1: Angle-Based Quantum Feature Encoding
             and Importance-Aware Weighting
Input:
    X ∈ ℝ^{N×D}              // Set of PCA-reduced feature vectors
    Clip                     // Clipping threshold
    w, v                     // Trainable feature scaling parameters
    Q = {q_1, q_1, …, q_D}   // Quantum register
Output:
    {|ψ⟩}                    //Encoded quantum states

    for each feature vector x in X do
        Clip x to the range [−clip, clip]
        Normalize x by clip and scale to [−π, π]
        for i = 1 to D do
            Apply R_y(x_i) on qubit q_i
            Apply R_y(w_i · x_i) on qubit q_i    // importance-
                                                    aware scaling
            Apply R_z(v_i · x_i) on qubit q_i
        end for
    end for
    return {|ψ⟩}
```

**Figure 7.** Pseudocode of the proposed algorithm, including the steps for normalizing input features via angle transformation, implementing angle-based quantum feature encoding, and applying the learnable importance-aware weighting mechanism.

### 2.3.1.1. Angle Transformation

Following the selection of ten T1Gd slices per patient via the Energy-Based Slice Selection method, each MRI image is subsequently processed using angle-based quantum feature encoding and the importance-aware weighting mechanism. The initial stage of this process, the angle transformation, involves resizing the images to a dimension of $img\_size \times img\_size$ using the resize function from the OpenCV library, followed by vectorization. In this study, the resized image dimension was set to 16 × 16, yielding a 256-dimensional feature vector for each slice. This image size was determined as optimal to preserve anatomically and contrastually discriminative information while maintaining a manageable qubit count to ensure compatibility with current NISQ constraints while minimizing computational overhead. Additionally, the INTER_AREA interpolation method was employed during the rescaling phase to minimize aliasing and artifacts during downsampling. Subsequently, pixel values of the vectorized slices were standardized using Z-score normalization to mitigate scale inconsistencies across different images. Z-score normalization quantifies the number of standard deviations a data point deviates from the mean of a distribution, enabling comparison across disparate distributions. The Z-score for a single data point is calculated as:

$$Z - score = \frac{(x - \mu)}{\sigma} \tag{3}$$

where $x$ represents the data point, and $\mu$ and $\sigma$ denote the mean and standard deviation of the distribution, respectively.

Given the limited number of qubits available in contemporary quantum computing hardware and simulators, dimensionality reduction is required prior to mapping the normalized GBM tumor features into the quantum space. In this study, Principal Component Analysis (PCA) was employed to control the number of qubits while retaining the most informative features with high variance. PCA aims to identify principal components—orthogonal axes representing directions of maximal variance in the data. In other words, it finds a linear transformation to a new coordinate system in which variance along each axis is maximized. Here, the number of principal components was automatically determined to capture 95% of the explained variance, while an upper limit was imposed to restrict quantum circuit complexity and qubit count. This operation is expressed as:

$$d_{components} = d_{qubits} = min(d_{PCA}, d_{MAX}) \tag{4}$$

where $d_{PCA}$ denotes the number of features obtained after PCA to achieve the 95% explained variance threshold, and $d_{MAX}$ is a predefined upper limit on the number of components. In this study, $d_{MAX}$ was set to an average of 18. Consequently, each resulting feature component is represented by a single qubit, defining the total number of qubits as $d_{qubits}$. Finally, to ensure compatibility with quantum gate parameters, the PCA-reduced features were linearly scaled and normalized to the range $[-\pi, \pi]$, completing the angle transformation stage.

*2.3.1.2. Angle-Based Quantum Encoding*

Following the definition of MRI features through the angle transformation stage, these features must be represented within the quantum space to enable efficient prediction of MGMT promoter methylation status. Quantum algorithms, computers, and simulators can only interact with and perform computations on data expressed as quantum states. The generation of quantum representations, referred to as the data encoding process, affects not only the preparation of the quantum algorithm but also the circuit design, often acting as a computational bottleneck [75,76]. For a detailed comparative analysis of the execution times of different feature encoding algorithms based on the number of data points and features, reference [77] can be consulted. Fundamentally, quantum feature encoding (embedding or mapping) is the process of projecting data defined in a classical vector space into a high-dimensional, nonlinear quantum Hilbert space where the inner product is defined [65,78,79]. This allows classical data to be expressed as quantum states [80–84]. In other words, by projecting data into a high-dimensional Hilbert space, the encoding process transforms relationships that are difficult to define directly in classical space into measurable and meaningful distance metrics within the quantum space [85]. Mathematically, given a labeled dataset defined in classical vector space as $\mathbb{D} = \left(x^d, y^d\right)_{d=1}^{D}$, only the feature vectors $x^d$ are mapped to quantum space, whereas the label information $y^d$ is retained in classical space. In this context, the quantum feature encoding process can be expressed as $\mathbb{D} \rightarrow \mathbb{W}$, where $\mathbb{W}$ is a subset of the complex vector space $\mathbb{C}^{N_q}$ representing the quantum data defined in Hilbert space, i.e., $\mathbb{W} \subset \mathbb{C}^{N_q}$.

Furthermore, for an $N$-dimensional feature vector, $x^d \in \mathbb{R}^N$ and $\mathbb{D} \subset \mathbb{R}^N$. For complex, heterogeneous, and highly correlated GBM tumor features, the Hilbert space provides noise resilience for MRI data obtained from different modalities, while offering high information capacity.

In the literature, numerous quantum feature encoding techniques have been proposed to project classical data into the broader feature space of quantum states (Hilbert space). Some methods offer advantages in terms of qubit efficiency, whereas others aim to reduce computational cost by shortening execution time. In this study, to maintain the qubit count within a specific range and improve the computational efficiency of the proposed specialized IA-QCNN architecture, the data were represented in Hilbert space using angle-based (rotational) quantum feature encoding. Angle-based encoding (also known as rotational encoding, qubit encoding, or tensor product encoding) is a fundamental technique used in quantum machine learning to project data from classical vector space to high-dimensional quantum Hilbert space where quantum states are defined. In this method, each initially defined classical datum is mapped to the rotation angles of single-qubit quantum gates [81,84,86–88]. Specifically, for a feature vector $x = [x_1, x_2,\ldots,x_N] \in R^N$, angle encoding represents each feature value through the rotation angles of single-qubit rotational gates, denoted as $R_x$, $R_y$ or $R_z$. The use of one qubit per feature with a constant-depth quantum circuit makes this method highly suitable for NISQ computers. Angle encoding is implemented by applying the tensor product of single-qubit rotation gates to the dataset and can be expressed as:

$$U(x) = \prod_{j=1}^{N} R_{ai}(x_i) \tag{5}$$

where $R_{ai}(.)$ denotes a rotation operation applied to the $i$-th qubit around the $ai \in \{X, Y, Z\}$ axis on the Bloch sphere. The single-qubit rotation gates around the $X, Y,$ and $Z$ axes are defined as:

$$R_x(x_i) = \begin{bmatrix} \cos\left(\frac{x_i}{2}\right) & -\sin\left(\frac{x_i}{2}\right)i \\ -\sin\left(\frac{x_i}{2}\right)i & \cos\left(\frac{x_i}{2}\right) \end{bmatrix}, R_y(x_i) = \begin{bmatrix} \cos\left(\frac{x_i}{2}\right) & -\sin\left(\frac{x_i}{2}\right) \\ \sin\left(\frac{x_i}{2}\right) & \cos\left(\frac{x_i}{2}\right) \end{bmatrix}, R_z(x_i) = \begin{bmatrix} e^{-i\frac{x_i}{2}} & 0 \\ 0 & e^{i\frac{x_i}{2}} \end{bmatrix}. \tag{6}$$

Through the angle encoding method, classical data directly influence the quantum state via trigonometric functions; thus, a hardware-efficient approach is achieved by maintaining a shallow circuit depth. Furthermore, since the number of qubits in the circuit is directly proportional to the number of defined features, a qubit-efficient representation is obtained, while the use of single-qubit quantum gates also provides notable temporal efficiency. Therefore, angle encoding is recommended in the literature for studies with a limited number of features, whereas for high-dimensional data, its use in combination with dimensionality reduction techniques is considered an efficient approach.

In this study, each feature component $x_i$, constrained to the $[-\pi, \pi]$ interval during the angle transformation stage, was mapped from classical space to the quantum Hilbert space using the $R_y$ rotational gate. This stage can be expressed as:

$$|\psi_i^1\rangle = R_y(x_i)|0\rangle \tag{7}$$

In the subsequent stage, an importance-aware weighting mechanism was applied to each feature to learnably scale its contribution to the quantum representation.

*2.3.1.3. Importance-Aware Weighting*

At this stage of the study, an importance-aware weighting approach is introduced, in which the contributions of features derived from the T1Gd pixel intensity information of GBM tumors to their quantum representations obtained via angle-based feature encoding are scaled through learnable parameters. In GBM tumors, which exhibit complex, heterogeneous, and highly correlated structures, certain features may contribute more prominently to the classification decision than others in predicting MGMT promoter methylation status directly from MR images. Accordingly, an approach is adopted in which clinically informative and discriminative tumor regions are emphasized by the proposed model through adaptive, differentiable, and learnable scaling parameters, while relatively less informative features are automatically suppressed during the training process. Within this framework, the contribution of each feature to its quantum representation is determined in a data-driven manner throughout training.

Furthermore, in contrast to conventional convolutional schemes that employ shared kernel weights based on spatial coordinates, the proposed method introduces an importance-aware weighting strategy by directly utilizing features derived from pixel intensity information, thereby extending the classical notion of "adaptive weighting" into the quantum computing domain. One of the key aspects to be emphasized here is that, rather than employing "if–else-based hard thresholding" approaches, a natural and soft thresholding behavior is learned through a gradient-based optimization process. In this way, the weighting mechanism is designed to be fully end-to-end learnable, without any external or rule-based intervention in the model's learning and decision-making process.

The proposed importance-aware weighting approach can be expressed in its most general form as:

$$\theta_i^{(eff)} = f(x_i; \alpha_i) \tag{8}$$

where $x_i$ denotes the $i$-th feature in the defined dataset, and $\alpha_i$ represents the learnable parameter that determines the contribution of this feature to its quantum representation. The function $f(.)$ corresponds to a linear (or nonlinear) scaling operation, which is optimized throughout the training process until optimal values are obtained.

In the proposed importance-aware weighting approach, each feature—initially represented as a quantum state via angle-based feature encoding, or equivalently expressed in terms of rotation angles using the $R_y$ rotational gate—is, at this stage, individually weighted through the $R_y$ and $R_z$ rotational gates and mapped into the quantum Hilbert space. In this way, instead of position-based fixed weights, a feature-dependent, learnable, and data-driven weighting operation is performed.

Importance-aware weighting using the $R_y$ rotational gate is expressed as:

$$|\psi_i^2\rangle = R_y(w_i x_i)|\psi_i^1\rangle \tag{9}$$

Importance-aware weighting using the $R_z$ rotational gate is expressed as:

$$|\psi_i^3\rangle = R_z(v_i x_i)|\psi_i^2\rangle \tag{10}$$

Here, in order to emphasize the relative contribution of each feature, the $R_y$ gate is applied in the first stage and the $R_z$ gate in the subsequent stage to the data that have been represented as quantum states via angle encoding. The parameters $w_i$ and $v_i$ are learnable coefficients that scale the $R_y$ transformation, which affects the amplitude distribution of the quantum state obtained via base angle encoding, and the $R_z$ transformation, which modifies the relative phase, respectively. Unlike classical hard-thresholding (if–else) mechanisms, the defined weighting parameters $w_i$ and $v_i$ provide a differentiable and soft importance-aware weighting criterion. Accordingly, during the training process, through the applied optimization step, features that contribute more significantly to the prediction of MGMT promoter methylation status are emphasized with larger effective rotation angles, whereas features containing less informative or noise-dominated content are suppressed with smaller rotation angles. In this context, the $R_y$ and $R_z$ rotational gates are integrated into the model with learnable parameters to represent the contributions of features to the amplitude and phase components of the quantum state, respectively.

By combining the above steps (Equations (9) and (10)), the final quantum state obtained for the $i$-th feature can be expressed as:

$$|\psi_i\rangle = R_z(v_i x_i) \, R_y(w_i x_i) |\psi_i^1\rangle \quad (11)$$

$$= R_z(v_i x_i) \, R_y(w_i x_i) \, R_y(x_i) |0\rangle$$

This expression defines how the proposed importance-aware weighting mechanism is realized within the framework of angle-based feature encoding. The circuit representation of this approach under a constrained number of qubits is illustrated in **Figure 8**.

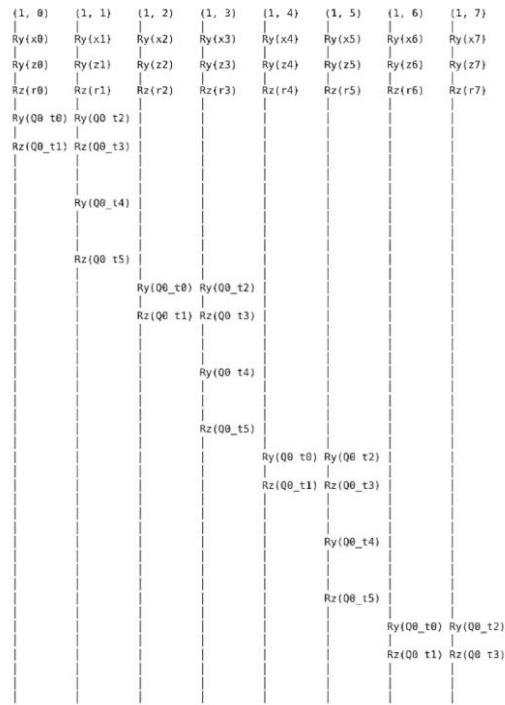

**Figure 8.** Quantum circuit diagram of the proposed angle-based quantum feature encoding and importance-aware weighting strategy, implemented using the TensorFlow Quantum (TFQ) and Cirq frameworks. For visual clarity, the circuit diagram is representatively illustrated using 8 qubits. In the actual experiments, the number of qubits was automatically determined based on the dimension obtained after PCA and constrained by the upper limit $d_{qubits} = min(d_{PCA}, d_{MAX})$ to restrict computational complexity. The full diagrams of the quantum circuits generated for all automatically determined qubit counts are presented in **Supplementary Figure S3**.

### 2.3.2. Ring-Topology Quantum Convolution and Folding-Based Quantum Pooling

In a Convolutional Neural Network (CNN) architecture, the convolution layer facilitates feature learning from input data through small, learnable filters; within this hierarchical structure, low-level features captured in the initial layers are progressively transformed into more abstract representations in deeper layers. The pooling layer reduces the spatial dimensions of feature maps, thereby decreasing computational overhead and mitigating the risk of overfitting. Building upon this fundamental framework, Quantum Convolutional Neural Network (QCNN) architectures—composed of sequentially applied quantum convolution and pooling layers—enable the extraction of increasingly large-scale features from data and have emerged as promising candidates for implementation on near-term quantum devices [67,79,89,90]. While preserving the core principles of classical CNNs, including measurement-induced nonlinearity, the locality of convolutional operations, and scalability to deep architectures, QCNNs exploit the high-dimensional Hilbert space of quantum systems to provide representations with enhanced expressive power for certain problem domains.

In this study, the proposed QCNN architecture enables the extraction of progressively larger-scale features from the T1Gd pixel intensity information of GBM patients through sequentially applied quantum convolution and pooling layers. Using these extracted feature representations, the prediction of MGMT promoter methylation status from MRI images is targeted. In the proposed QCNN structure, the quantum convolution layer consists of parameterized single-qubit quantum gates arranged in a nested manner and two-qubit quantum blocks that generate entanglement. More specifically, the quantum convolution layer is constructed by applying parametric two-qubit blocks—composed of $R_y$ and $R_z$ rotation gates defined with shared parameters over neighboring qubit pairs, along with $CNOT$ gates used to establish entanglement between qubit pairs—according to a ring topology. The use of shared parameters not only reduces the number of trainable parameters but also ensures that the same feature patterns can be recognized across different regions of the circuit. The entanglement structure formed by $CNOT$ two-qubit quantum gates is referred to as "cyclically repeated entanglement" or "ring entanglement." Through this structure, each qubit interacts not only with its immediate neighbor but also with its shifted neighbors, enabling the learning of local correlations and the propagation of this information throughout the circuit. The mathematical structure of the proposed quantum convolution layer over a pair of qubits can be expressed as follows:

$$U_{conv}(q_1, q_2, \theta) = R_z^{(q_2)}(t_5).CNOT_{q_1 \to q_2} R_y^{(q_2)}(t_4).CNOT_{q_1 \to q_2}.\left(R_z^{(q_2)}(t_3)R_y^{(q_2)}(t_2)\right).\left(R_z^{(q_1)}(t_1)R_y^{(q_1)}(t_0)\right). \quad (11)$$

Here, $\theta = [t_0, t_1, t_2, t_3, t_4, t_5]^T$ defines the trainable parameters shared between qubit pairs throughout the entire circuit. In other words, the two-qubit convolutional block, denoted as $U_{conv}(q_1, q_2, \theta)$, is applied to all neighboring qubit pairs using the same parameter set. This structure can be interpreted as the quantum analog of the weight-sharing principle employed in classical convolutional neural networks. Moreover, the unitary entanglement block within the quantum convolution layer, responsible for propagating correlation information across the circuit, is expressed as:

$$U_{ent}(q_1, q_2, \theta) = CNOT_{q_1 \to q_2} R_y^{(q_2)}(t_4).CNOT_{q_1 \to q_2}. \quad (12)$$

In this formulation, $q_1$ represents the control qubit and $q_2$ represents the target qubit. This configuration allows the degree of entanglement between the two qubits to be learned in an end-to-end and dataset-specific manner via the trainable parameter $t_4$. Consequently, the final form of the quantum convolution layer defined for a qubit pair is expressed as:

$$U_{conv}(q_1, q_2, \theta) = R_z^{(q_2)}(t_5).U_{ent}.\left(R_z^{(q_2)}(t_3)R_y^{(q_2)}(t_2)\right).\left(R_z^{(q_1)}(t_1)R_y^{(q_1)}(t_0)\right) \quad (13)$$

The proposed quantum convolution block is implemented across all qubit pairs in accordance with a ring topology. This structure is achieved by applying the convolution block in two stages under periodic boundary conditions. In this architecture, where both stages are executed in parallel within the circuit, the ring neighborhood sets are defined as:

$$\varepsilon_{even} = \{(0,1),(2,3),\ldots,(N-2,N-1)\}$$

$$\varepsilon_{odd} = \{(1,2),(3,4),\ldots,(N-1,0)\}$$
(14)

Under these conditions, the quantum convolution block based on the ring topology is formulated as follows:

$$U_{ring\_conv}(q_1,q_2,\theta) = \left(\prod_{(q_1,q_2)\in \varepsilon_{odd}} U_{conv}(q_1,q_2,\theta)\right)\left(\prod_{(q_1,q_2)\in \varepsilon_{even}} U_{conv}(q_1,q_2,\theta)\right)$$
(15)

A crucial point to emphasize is that the $(N-1,0)$ pairing ensures the periodic boundary condition by returning to the initial qubit state, thereby establishing a complete ring topology. This design not only enhances the parallel processing capacity on hardware but also ensures the establishment of a global correlation among all neighboring qubits. When the $U_{ring\_conv}$ structure defined for two qubits is generalized for all qubits in the circuit, the evolution of the wavefunction at the $l$-th layer can be expressed as:

$$\left|\psi_i^{(0)}\right\rangle = R_z(v_i x_i)\, R_y(w_i x_i)\, R_y(x_i)|0\rangle$$

$$\left|\Psi^{(l-1)}\right\rangle = \otimes_{i=1}^d \left|\psi_i^{(0)}\right\rangle$$

$$\left|\Psi^{(l)}\right\rangle = U_{ring\_conv}^{(l)}(\theta^l)\left|\Psi^{(l-1)}\right\rangle$$

$$\left|\Psi^{(l)}\right\rangle = \left(\prod_{(i,j)\in \varepsilon_{odd}} U_{conv}^{(i,j)}(\theta^l)\right)\left(\prod_{(k,m)\in \varepsilon_{even}} U_{conv}^{(k,m)}(\theta^l)\right)\left|\Psi^{(l-1)}\right\rangle$$
(16)

Following the quantum convolution layer, a quantum pooling operation is applied to the high-dimensional feature maps—represented as quantum states and derived from the T1Gd pixel intensity data of GBM patients. The primary objectives of this operation are to reduce the qubit count, establish a hierarchical representation, and enhance computational efficiency. Similar to the pooling layers in classical CNN architectures, the quantum pooling operation yields a more compact representation by reducing the system's degrees of freedom at each layer. Various approaches have been proposed in the literature to implement quantum pooling, primarily categorized into measurement-based pooling and unitary transformation-based pooling. In measurement-based pooling, a qubit designated for removal is measured, and the measurement outcome controls a unitary transformation applied to a neighboring qubit. In unitary transformation-based pooling, parametrized quantum gates are applied to two qubits, and by taking the partial trace over one qubit, the two-qubit state is reduced to a single-qubit state, allowing information to propagate through the remaining qubit across the circuit.

In this study, to avoid the mid-circuit measurement limitations inherent to current NISQ devices, quantum pooling was implemented via unitary transformations. Following the quantum convolution operation, qubits were partitioned into source (passive) and target (active) sets. Using the "folding-based strategy" during the pooling process, the first half of the qubits was projected onto the second half through unitary operations. The quantum pooling operation defined between two qubits, along with its associated unitary operator, can be formally expressed as:

$$U_{pool}(q_s, q_t, \phi) = R_y^{(q_t)}(p_2) \cdot CNOT_{q_s \to q_t} \cdot R_z^{(q_s)}(p_1) \cdot R_y^{(q_t)}(p_0) \tag{17}$$

Here, $q_s$ and $q_t$ denote the source and target qubits, respectively. $\phi = [p_0, p_1, p_2]^T$ defines the trainable parameters shared between each qubit pair during the pooling operation. To extend this quantum pooling operation across all qubits in the circuit, the unitary operator can be generalized as follows:

First, the total number of qubits $N$ is divided into two subsets, source ($S$) and target ($T$), in accordance with the folding-based strategy. In other words, the first half of the data is assigned to the $S$ subset, while the remaining half is assigned to the $T$ subset. This can be expressed as:

$$S^{(l)} = \{q_0, \ldots, q_{N/2-1}\}, \qquad T^{(l)} = \{q_{N/2}, \ldots, q_{N-1}\} \tag{18}$$

Here, the set $S^{(l)}$ represents the source qubits, whereas $T^{(l)}$ corresponds to the target qubits onto which the information will be projected. The $U_{pool}$ block is applied between the $i$-th source qubit and the $(i + N/2)$-th target qubit within the $l$-th layer. Accordingly, the resulting quantum pooling layer is expressed as:

$$U_{pool\_layer}^{(l)}(\phi^l) = \prod_{i=0}^{N/2-1} U_{pool}(q_i, q_{i+N/2}, \phi^l)$$

$$U_{pool\_layer}^{(l)}(\phi^l) = \prod_{i=0}^{N/2-1} U_{pool}(q_i \in S^{(l)}, q_{i+N/2} \in T^{(l)}, \phi^l) \tag{19}$$

$$U_{pool\_layer}^{(l)}(\phi^l) = \prod_{i=0}^{N/2-1} \left( R_y^{(q_{i+N/2})}(p_2^l) \cdot CNOT_{q_i \to q_{i+N/2}} \cdot R_z^{(q_i)}(p_1^l) \cdot R_y^{(q_{i+N/2})}(p_0^l) \right)$$

When the quantum convolution and pooling layers are combined, the evolution of the circuit's wavefunction at the end of each $l$-th layer can be expressed as:

$$|\Psi^{(l)}\rangle = U_{pool\_layer}^{(l)}(\phi^l) \, U_{ring\_conv}^{(l)}(\theta^l) |\Psi^{(l-1)}\rangle$$

$$|\Psi^{(l)}\rangle = \left[ \prod_{k=0}^{N/2-1} \left( R_y^{(q_{k+N/2})}(p_2^l) \cdot CNOT_{q_k \to q_{k+N/2}} \cdot R_z^{(q_k)}(p_1^l) \cdot R_y^{(q_{k+N/2})}(p_0^l) \right) \right] \tag{20}$$

$$\left[ \left( \prod_{(i,j) \in \varepsilon_{odd}} U_{conv}^{(i,j)}(\theta^l) \right) \left( \prod_{(m,n) \in \varepsilon_{even}} U_{conv}^{(m,n)}(\theta^l) \right) \right] |\Psi^{(l-1)}\rangle$$

In this framework, while the $U_{ring\_conv}^{(l)}(\theta^l)$ layer generates a robust representation by learning local features from the data, the $U_{pool\_layer}^{(l)}(\phi^l)$ layer reduces the qubit count, mapping these features into a lower-dimensional Hilbert space to establish a hierarchical representation. Accordingly, as a result of all transformations undergone by the wavefunction across $L$ layers, the final wavefunction is expressed as:

$$|\Psi_{final}\rangle = \prod_{l=1}^{L} U_{pool\_layer}^{(l)}(\phi^l) \, U_{ring\_conv}^{(l)}(\theta^l) |\Psi^{(0)}\rangle \tag{21}$$

The algorithmic structure of the proposed ring-topology quantum convolution and folding-based quantum pooling approaches is presented in detail in **Supplementary Figure S4**, and a circuit representation with a limited number of qubits is provided in **Figure 9**.

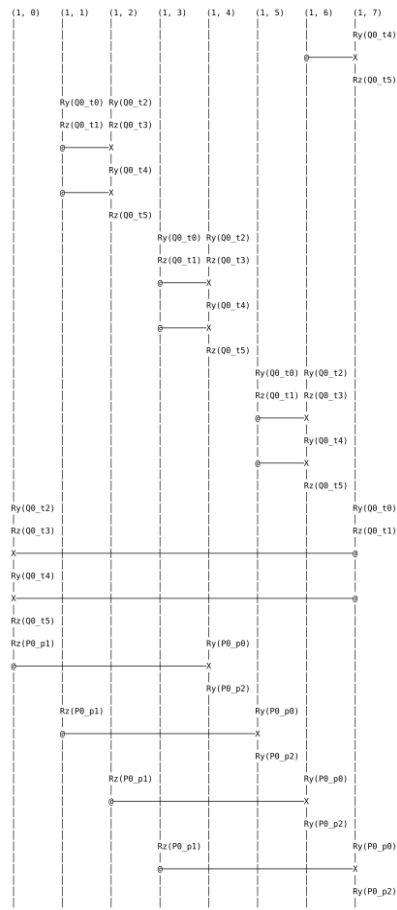

**Figure 9.** Quantum circuit diagram of the proposed ring-topology quantum convolution and folding-based quantum pooling approaches, implemented using the TensorFlow Quantum (TFQ) and Cirq frameworks. Due to space constraints, the circuit diagram is representatively illustrated for 8 qubits. The complete diagrams for the quantum circuits generated for all automatically determined qubit counts are provided in **Supplementary Figure S5**.

### 2.3.3. Measurement

In the preceding step, high-dimensional quantum feature maps were generated from the T1Gd pixel intensity data of GBM patients using the proposed ring-topology quantum convolution layer, followed by the construction of a robust hierarchical representation through the folding-based quantum pooling layer. In this stage, to predict the MGMT promoter methylation status from MRI images based on these feature representations, a measurement operation is performed on the hierarchically reduced target qubit set, $T^{(l)}$, via the Pauli-Z ($\hat{Z}$) operator. Specifically, the expectation value for each target qubit is calculated based on the final wavefunction of the quantum circuit, $|\Psi_{final}\rangle$, and returned as a list to obtain the probability distribution for each class. This process is expressed as follows:

$$\hat{y}_i = \langle \Psi_{final} | \hat{Z}_i | \Psi_{final} \rangle, \qquad i \in \{0, 1, \ldots, N_{final} - 1\} \tag{22}$$

where $N_{final}$ represents the number of active qubits remaining after the final pooling layer. A crucial point to emphasize here is that, as the circuit depth increases and the number of active qubits decreases due to the hierarchical pooling process, a fixed-size input is consistently provided to the Softmax function. Accordingly, the Pauli-Z expectation values are calculated over the first four active qubits remaining after pooling, thereby maintaining the necessary representational power of the circuit through this fixed dimension.

Furthermore, since the positive (methylated) and negative (unmethylated) class labels representing the MGMT promoter methylation status of GBM patients were already encoded by the dataset providers, no additional processing was required at this step. However, as ten T1Gd slices were determined for each patient using the Energy-Based Slice Selection method, the class labels were expanded from a patient-based to a slice-based format. The class label for each patient was repeated ten times—corresponding to the number of selected slices—to ensure precise alignment between the slice data and the patient labels.

### 2.3.4. Softmax function, loss calculation and optimization

In this stage, the measurement results in the Pauli-Z basis, obtained from the first four qubits of the hierarchically reduced target qubit set, are first returned as a list. Each element in this list represents the classical value derived from the measurement of the corresponding qubit. Subsequently, these quantum expectation values are fed as input into a classical Dense (or fully connected) layer with two neurons to learn the discriminative features of the methylated and unmethylated classes. The raw scores obtained from the Dense layer are then passed through a Softmax function to compute the class probabilities. This process is expressed as follows:

$$z_j = \sum_{i=0}^{3}(w_{ji} \cdot \hat{y}_i) + b_j, \quad j \in \{0,1\}$$

$$p_j = \frac{exp(z_j)}{\sum_{k=0}^{1} exp(z_k)} \tag{23}$$

Here, $w_{ji}$ and $b_j$ are the weight and bias parameters, respectively, which are adjustable and learnable through optimization. The index j represents the methylated and unmethylated classes, while $p_j$ denotes the class probabilities obtained after the Softmax function. In the following step, to minimize the discrepancy between the predicted probabilities ($p_j$) and the ground truth labels ($y_j$), Sparse Categorical Cross-Entropy was employed as the Loss Function:

$$L = -\sum_{j=0}^{1} y_j \, log(p_j) \tag{24}$$

In this study, the Adam (Adaptive Moment Estimation) optimization algorithm was utilized to minimize the loss function and update the parameters in both the quantum and classical layers.

### 2.3.5. Evaluation Metrics

In this study, the classification performance of the proposed specialized IA-QCNN architecture in predicting MGMT promoter methylation status from MRI images of GBM tumors was evaluated using metrics including average training loss, validation loss, training accuracy, and validation accuracy. Performance variations corresponding to each training epoch are presented through graphical plots. Furthermore, to provide a more detailed analysis of the model's ability to discriminate between methylated and unmethylated classes, fundamental metrics such as precision, recall, and the F1-score were computed for each class individually.

## 3. Results

In this section, the proposed specialized IA-QCNN architecture was evaluated on the RSNA-MICCAI Brain Tumor Radiogenomic dataset for predicting MGMT promoter methylation status from MRI images of GBM tumors, and the obtained results are presented in detail. This section is organized into six subsections, based on the analyzed parameters and experimental scenarios:

- Results of the specialized IA-QCNN architecture in predicting MGMT promoter methylation from T1Gd MRI images,
- Results of the specialized IA-QCNN architecture in predicting MGMT promoter methylation from mpMRI images,
- Results of the specialized IA-QCNN architecture under mild-to-moderate Gaussian noise in T1Gd MRI images,
- Results of the specialized IA-QCNN architecture under gate-level noise in the quantum circuit,
- Results of the specialized IA-QCNN architecture under hybrid noise (multi-level Gaussian perturbations) in both T1Gd MRI images and quantum gates,
- Comparative results of CNN and DNN models designed with a similar number of parameters, along with Transfer Learning (TL) and State-of-the-Art (SOTA) DL models.

In this study, the initial learning rate for the Adam optimizer was set to $3x10^{-3}$, and was subsequently reduced automatically based on the model's learning progress using the ReduceLROnPlateau module from the Keras library, with a minimum learning rate of $2x10^{-4}$. The number of epochs was initially set to 60 with a batch size of 16; additionally, an Early Stopping module was incorporated to mitigate overfitting during the training process. For each iteration, the average training loss, validation loss, training accuracy, and validation accuracy were recorded. To maintain computational efficiency and reduce processing time, the circuit depth ($L$) was set to 1. The number of qubits was automatically determined based on the dimensionality obtained after PCA and constrained by an upper bound defined as $d_{qubits} = min(d_{PCA}, d_{MAX})$ to limit computational complexity.

Quantum circuit operations were executed using the TensorFlow Quantum (TFQ) 0.7.6 library, performing state-vector simulations on a Cirq 1.5.0 backend. The classical DL architecture was implemented using the Keras library integrated with TensorFlow 2.18.1. High-performance quantum simulations and model training were conducted on the Perlmutter supercomputer at NERSC, while process management and data preparation were carried out on a local system equipped with a 12th Generation Intel(R) Core(TM) i5 processor and 8.0 GB of RAM.

## 3.1. Results of the specialized IA-QCNN architecture in predicting MGMT promoter methylation from T1Gd MRI images

In this section, the prediction of MGMT promoter methylation status in GBM tumors was performed using the specialized Importance-Aware Quantum Convolutional Neural Network with Ring-Topology (IA-QCNN) architecture from T1Gd MRI images, effectively distinguishing between methylated (positive) and unmethylated (negative) cases. During this classification process, the performance of the specialized IA-QCNN architecture was analyzed based on accuracy and loss values obtained throughout the training and validation phases. The training process of the IA-QCNN architecture was conducted on a slice-wise basis, and the procedure was terminated at the 17th iteration via the Early Stopping mechanism. In this context, the average training and validation accuracies were calculated as 0.56 and 0.62, respectively, while the corresponding training and validation losses were recorded as 0.69 and 0.67. The accuracy and loss curves corresponding to these values are presented in **Figure 10**.

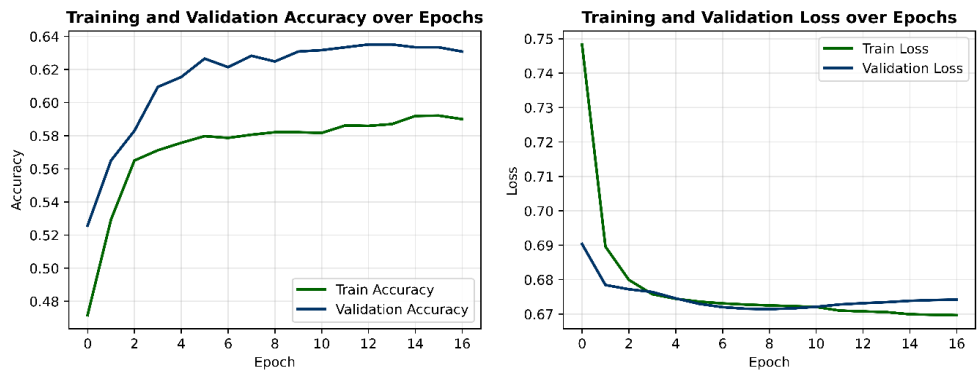

**Figure 10.** Epoch-wise variation of slice-level accuracy and loss values for the prediction of MGMT promoter methylation status in GBM tumors using T1Gd MRI images via the IA-QCNN architecture.

In the subsequent stage, the slice-based prediction probabilities obtained for each patient were combined using a "patient-level approach", in alignment with the clinical decision-making process. At this stage, the probability values of the slices belonging to each patient were integrated using the mean aggregation method. Subsequently, the class with the highest probability value was selected to generate the patient-level prediction. As a result of this hierarchical evaluation, the overall patient-level accuracy of the model on the test set was recorded as 0.67. The patient-level confusion matrix for the prediction of MGMT promoter methylation status via the IA-QCNN architecture is presented in **Figure 11**, and the ROC curve is provided in **Figure 12**.

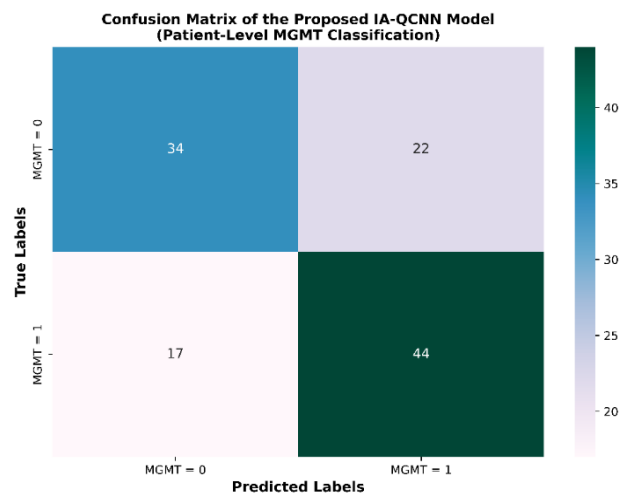

**Figure 11.** Confusion matrix generated by averaging slice-level IA-QCNN predictions derived from T1Gd MRI images for patient-level prediction of MGMT promoter methylation status in GBM tumors.

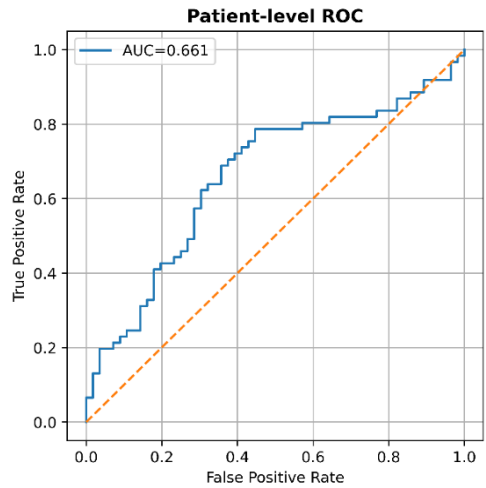

**Figure 12.** Receiver Operating Characteristic (ROC) curve obtained by aggregating slice-level IA-QCNN predictions derived from T1Gd MRI images into a patient-level representation for predicting MGMT promoter methylation status in GBM tumors. An AUC score of 0.66 was achieved.

As a result of aggregating slice-level predictions into patient-level representations via mean averaging for each patient, the precision, recall, and F1-score values for the MGMT promoter methylated (positive) class were obtained as 0.67, 0.72, and 0.69, respectively; for the unmethylated class, these values were 0.67, 0.61, and 0.64, respectively. The detailed graphical representation of the class-wise precision, recall, and F1-score values is presented in **Figure 13**.

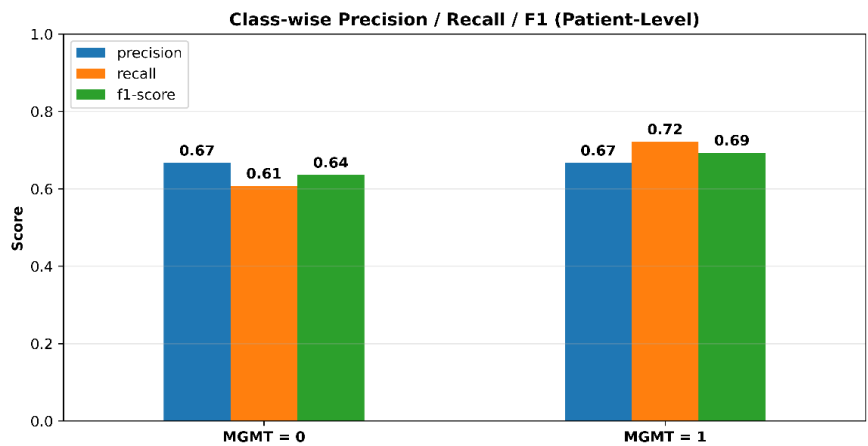

**Figure 13.** Detailed representation of precision, recall, and F1-score values for methylated (1) and unmethylated (0) classes according to MGMT promoter methylation status.

### 3.2. Results of the specialized IA-QCNN architecture in predicting MGMT promoter methylation from mpMRI images

At this stage of the study, the prediction of MGMT promoter methylation status in GBM tumors was performed using the proposed IA-QCNN architecture based on multi-parametric (mp) MRI modalities, including FLAIR, T1-weighted (T1w), T1Gd, and T2 images, thereby distinguishing between methylated and unmethylated cases. For each patient, ten T1Gd slices were initially determined using the energy-based slice selection method. Subsequently, the corresponding T1w, T2, and FLAIR modalities for these selected slice indices were obtained, and a single fused image representation was generated by computing the pixel-wise average of these four modalities. When these fused mpMRI feature maps were utilized for training the proposed IA-QCNN architecture, the procedure was terminated at the 12th iteration via the Early Stopping mechanism. Following the slice-based training process, the mean training and validation accuracies were calculated as 0.59 and 0.46, respectively, while the training and validation losses were recorded as 0.67 and 0.74. The epoch-wise variations of the accuracy and loss curves corresponding to these performance metrics are presented in **Figure 14.**

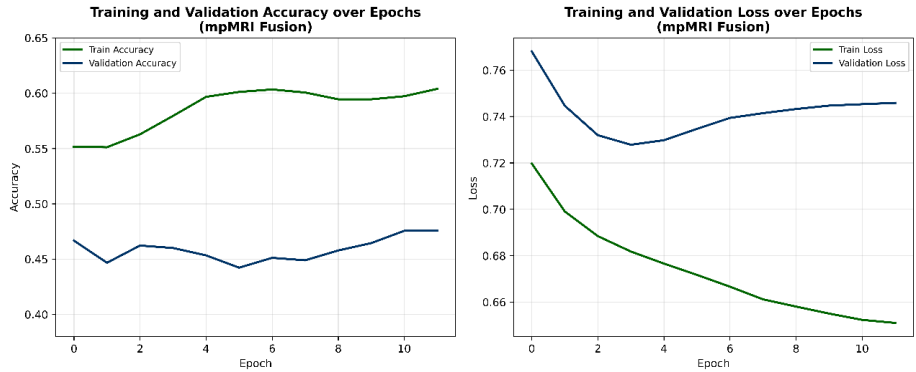

**Figure 14.** Epoch-wise variation of slice-level accuracy and loss values for the prediction of MGMT promoter methylation status in GBM tumors using mpMRI images via the IA-QCNN architecture.

As implemented for the T1Gd MRI modality in the previous stage, a mean aggregation method was employed here to transition from slice-level results to a patient-level approach. Following the testing phase, the overall patient-level accuracy on the mpMRI test set was recorded as 0.49. The patient-level confusion matrix, illustrating the performance of the proposed IA-QCNN architecture in predicting MGMT promoter methylation status using mpMRI data, is presented in **Figure 15**, while the ROC curve, documenting the model's discriminative power, is provided in **Figure 16**.

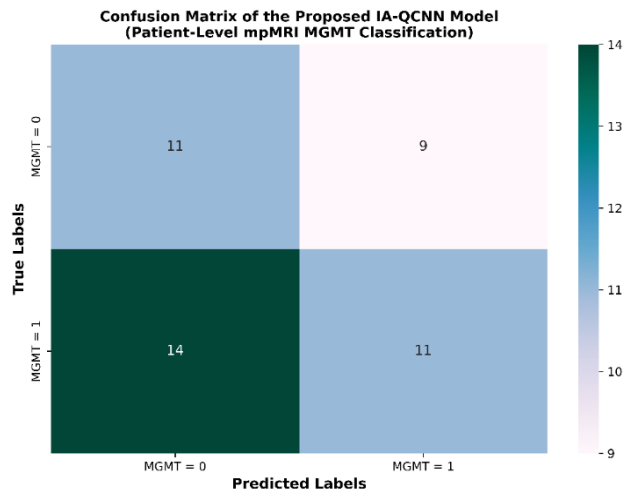

**Figure 15.** Patient-level confusion matrix for the prediction of MGMT promoter methylation status in GBM tumors derived from mpMRI images.

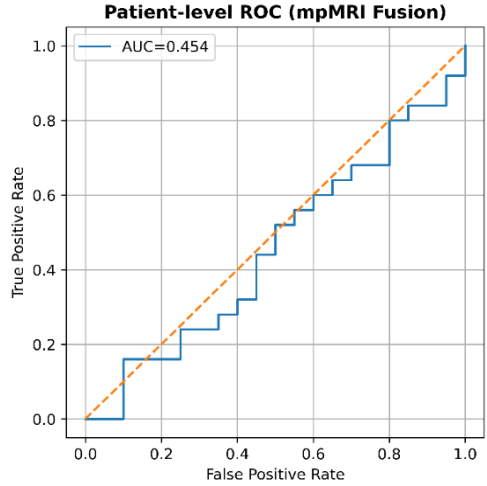

**Figure 16.** ROC curve and corresponding AUC value (AUC = 0.45) obtained by aggregating slice-level IA-QCNN predictions into a patient-level representation for MGMT promoter methylation in GBM tumors using mpMRI images.

As a result of the patient-level mean aggregation of slice-based predictions performed with mpMRI modalities for each patient, the precision, recall, and F1-score values for the methylated class were obtained as 0.55, 0.44, and 0.49, respectively. For the unmethylated class, these values were recorded as 0.44, 0.55, and 0.49, respectively. The detailed graphical representation of these class-wise precision, recall, and F1-score values is provided in **Figure 17**.

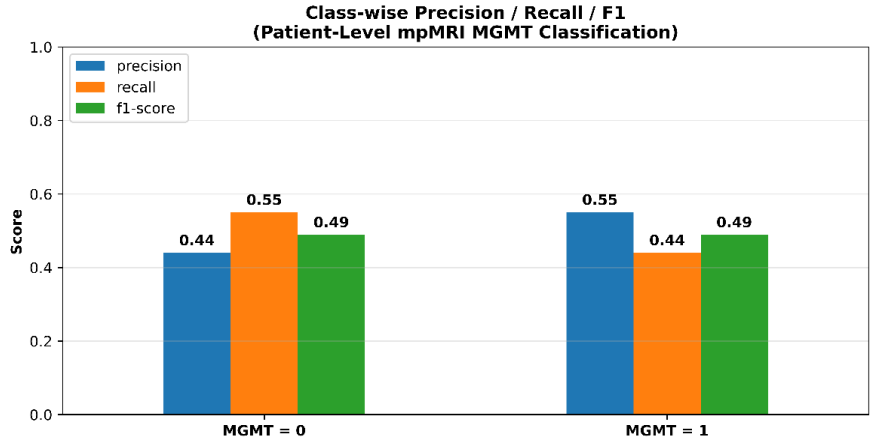

**Figure 17.** Detailed representation of class-wise precision, recall, and F1-score values obtained by aggregating slice-level predictions of the specialized IA-QCNN model into patient-level representation using mpMRI modalities.

### 3.3. Results of the specialized IA-QCNN architecture under mild-to-moderate Gaussian noise in T1Gd MRI images

In this phase of the study, Gaussian noise was added to all mpMRI images of GBM tumors to evaluate the real-world applicability, generalization capability, reliability, and robustness of the proposed IA-QCNN architecture. Specifically, random values sampled from a Gaussian distribution with zero mean and a standard deviation ($\sigma$) of 0.03 were added to all normalized mpMRI pixel intensities. In other words, Gaussian noise was implemented at the image level by introducing zero-mean Gaussian perturbations with a standard deviation of 0.03 to the normalized intensity values on a pixel-wise basis. Following the noise injection process, pixel values were rescaled to the [0, 1] range. Thus, the model's robustness against real-world scenarios—such as variability in scanner protocols, operator-induced inconsistencies, and heterogeneity in imaging parameters—was comprehensively evaluated. In the subsequent stage, ten T1Gd slices for each patient were identified from the Gaussian-noise-perturbed dataset using the energy-based slice selection method. When these noise-affected T1Gd MRI images were utilized for training the proposed IA-QCNN architecture, the procedure was terminated at the 17th iteration via the Early Stopping mechanism. Under these conditions, the mean training accuracy, validation accuracy, training loss, and validation loss were recorded as 0.56, 0.58, 0.68, and 0.68, respectively. The epoch-wise variation of the accuracy and loss values during the training and validation phases is presented in **Figure 18**.

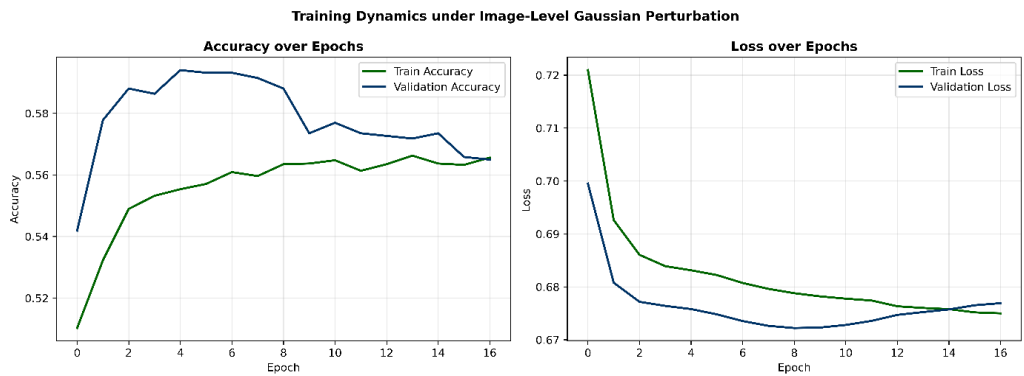

**Figure 18.** Epoch-wise variation of slice-level training and validation accuracy/loss values for the IA-QCNN architecture in predicting MGMT promoter methylation status under the addition of Gaussian noise to T1Gd MRI images.

Additionally, for the Gaussian-noise-perturbed T1Gd MRI images, a mean aggregation method was employed to transition from slice-level results to a patient-level approach. As a result of this hierarchical evaluation, the overall patient-level accuracy of the model on the noisy test set was recorded as 0.61. The patient-level confusion matrix illustrating the performance of the IA-QCNN architecture in predicting MGMT promoter methylation status under noisy conditions is presented in **Figure 19**, while the ROC curve is provided in **Figure 20**.

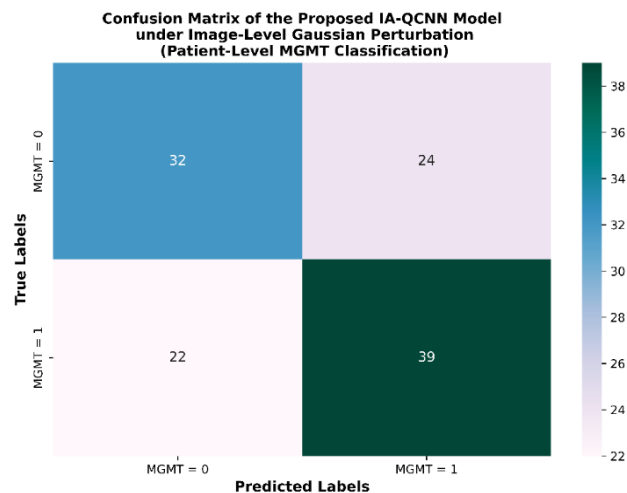

**Figure 19.** Patient-level confusion matrix of the IA-QCNN architecture evaluated on Gaussian-noise-perturbed T1Gd MRI images for predicting MGMT promoter methylation status.

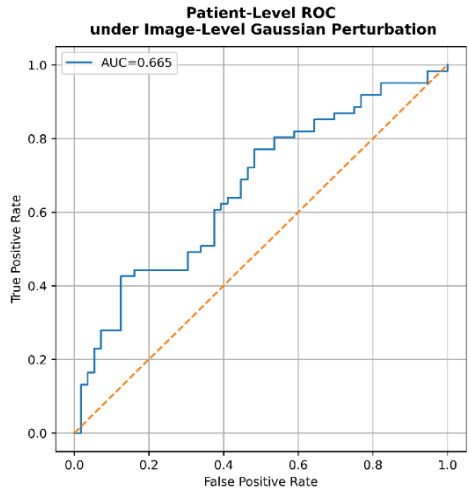

**Figure 20.** ROC curve and corresponding AUC value (AUC = 0.67) obtained by aggregating slice-level predictions into patient-level assessments for the IA-QCNN architecture evaluated on Gaussian-noise-perturbed T1Gd MRI images.

As a result of aggregating slice-level predictions obtained from Gaussian-noise-perturbed T1Gd MRI modalities into patient-level representations via mean averaging for each patient, the precision, recall, and F1-score values for the MGMT promoter methylated class were obtained as 0.62, 0.64, and 0.63, respectively. For the unmethylated class, these values were 0.59, 0.57 and 0.58, respectively. The detailed distribution of these class-wise metrics is presented in **Figure 21**.

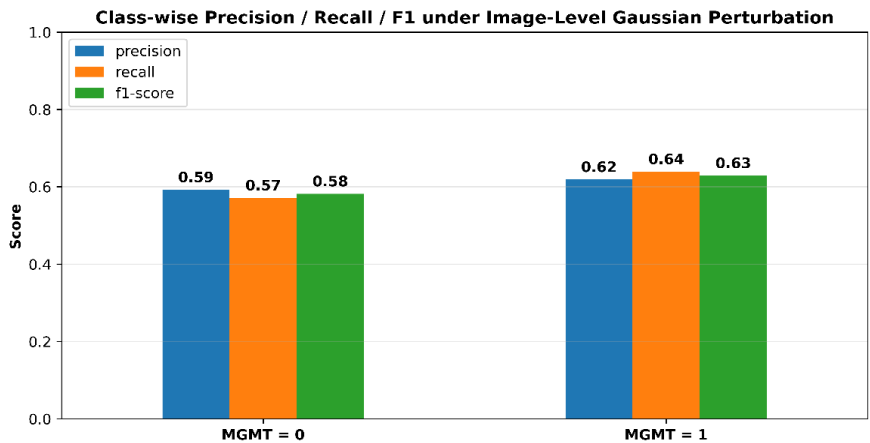

**Figure 21.** Detailed representation of class-wise precision, recall, and F1-score values, obtained by the patient-level mean aggregation of slice-level predictions from the specialized IA-QCNN model using Gaussian-noise-perturbed T1Gd MRI modalities.

### 3.4. Results of the specialized IA-QCNN architecture under gate-level noise in the quantum circuit

In this section, Gaussian noise was introduced to the parameters of the quantum circuit to approximately model gate-induced errors occurring in real-world quantum computing environments while maintaining computational efficiency. In this manner, the effect of gate-level noise was modeled to a certain extent within the quantum simulation framework, and its impact on the classification performance of the specialized IA-QCNN architecture was investigated. Gaussian noise was applied to the symbolic parameter values prior to circuit execution, thereby perturbing only the rotation angles without altering the overall circuit architecture. Specifically, Gaussian noise was added to all rotational gates within the angle-based quantum feature encoding, importance-aware weighting, ring-topology quantum convolution, and folding-based quantum pooling stages. This approach affected the rotation angles propagated throughout the circuit without altering the fundamental circuit design. In this stage, random values derived from a Gaussian distribution with a mean of 0 and a standard deviation of 0.02 were added to the rotational gates. When the IA-QCNN model was trained with Gaussian-noise-perturbed gates, the training process was terminated at the 18th iteration via the Early Stopping mechanism. Under these conditions, the mean training accuracy, validation accuracy, training loss, and validation loss were recorded as 0.58, 0.56, 0.67, and 0.68, respectively. The variation in accuracy and loss values during the training and validation phases is illustrated in **Figure 22**.

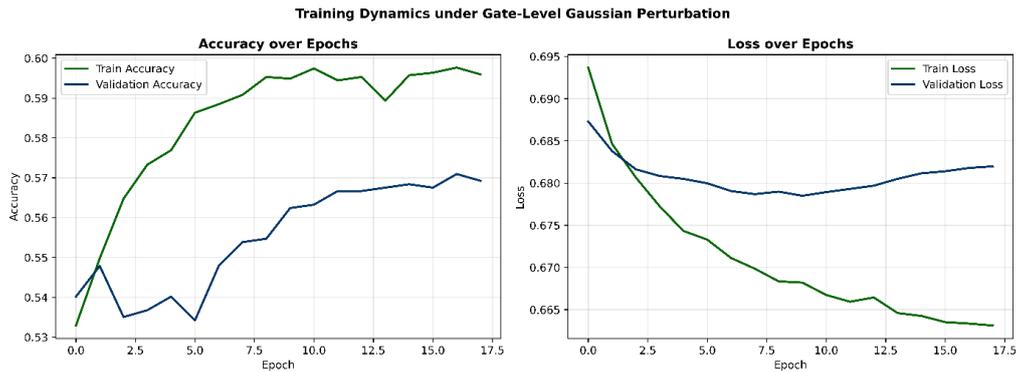

**Figure 22.** Epoch-wise variation of slice-level training and validation accuracy and loss values for the IA-QCNN architecture in predicting MGMT promoter methylation status under the addition of Gaussian noise to the rotational gate parameters within the quantum circuit.

Furthermore, when transitioning from slice-level results to patient-level outcomes, the overall patient-level accuracy of the IA-QCNN model with Gaussian-noise-perturbed gate parameters was obtained as 0.51 on the test set. The patient-level confusion matrix, demonstrating the performance of the IA-QCNN architecture in predicting MGMT promoter methylation status under these conditions, is presented in **Figure 23**, and the ROC curve is provided in **Figure 24**.

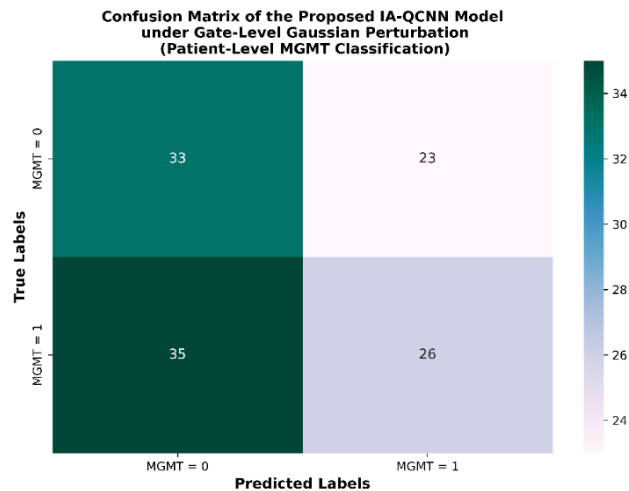

**Figure 23.** Patient-level confusion matrix for the prediction of MGMT promoter methylation status via the IA-QCNN architecture under the addition of Gaussian noise to the rotational gate parameters in the quantum circuit.

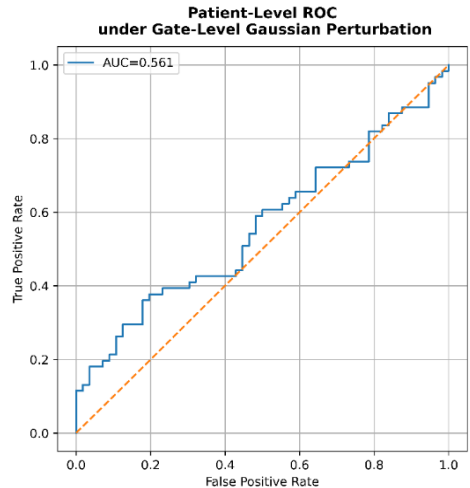

**Figure 24.** ROC curve and AUC value (AUC = 0.56) for the IA-QCNN architecture under the addition of Gaussian noise to the rotational gate parameters in the quantum circuit, obtained by aggregating slice-level predictions on a patient-level basis.

Upon aggregating slice-level predictions of the specialized IA-QCNN architecture into patient-level representations via averaging under Gaussian noise applied to the rotational gate parameters, the precision, recall, and F1-score for the methylated class were obtained as 0.53, 0.43, and 0.47, respectively. For the unmethylated class, these values were 0.49, 0.59, and 0.53, respectively. The detailed distribution of these class-wise metrics is presented in **Figure 25**.

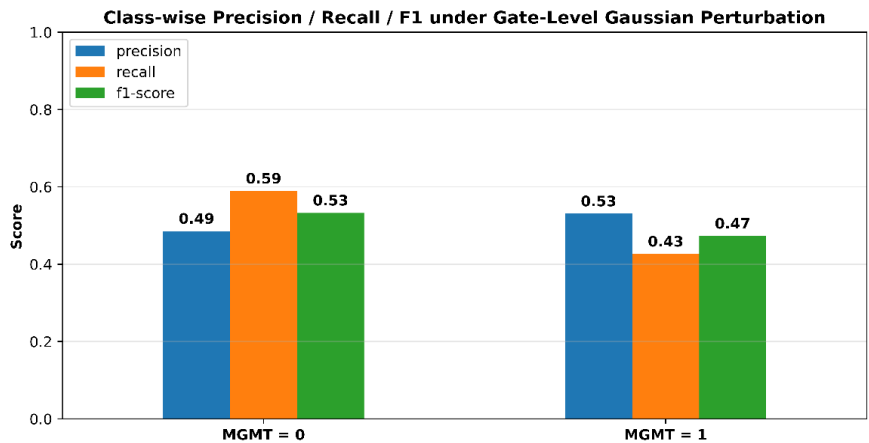

**Figure 25.** Detailed representation of class-wise precision, recall, and F1-score values, obtained by the patient-level mean aggregation of slice-level predictions from the customized IA-QCNN model under the addition of Gaussian noise to the rotational gate parameters within the quantum circuit.

### 3.5. Results of the specialized IA-QCNN architecture under hybrid noise (multi-level Gaussian perturbations) in both T1Gd MRI images and quantum gates

In this final stage, where the proposed specialized IA-QCNN architecture was evaluated under different noise scenarios, a hybrid noise environment was constructed to assess the model's ability to distinguish between methylated and unmethylated classes. In this process, a parallel noise mechanism was designed, and the classification performance was analyzed with Gaussian noise applied in parallel to both T1Gd MRI images and rotational gate blocks within the quantum circuit. In this way, the model's suitability for real-world conditions and its generalization capability were evaluated, while its effectiveness under environments that approximately simulate gate-level errors occurring in real quantum computers was also investigated. In this stage, Gaussian noise with a mean of 0 and a standard deviation of 0.03 was added to the T1Gd MRI data, while Gaussian noise with a mean of 0 and a standard deviation of 0.02 was applied to the rotational gates. When the IA-QCNN model was trained using Gaussian-noise-perturbed T1Gd MRI data and Gaussian-noise-perturbed quantum gates, the training process was terminated at the 16th iteration via the Early Stopping mechanism. Under these conditions, the mean training accuracy, validation accuracy, training loss, and validation loss were 0.57, 0.64, 0.68, and 0.65, respectively. The variation of accuracy and loss during the training and validation phases is presented in **Figure 26.**

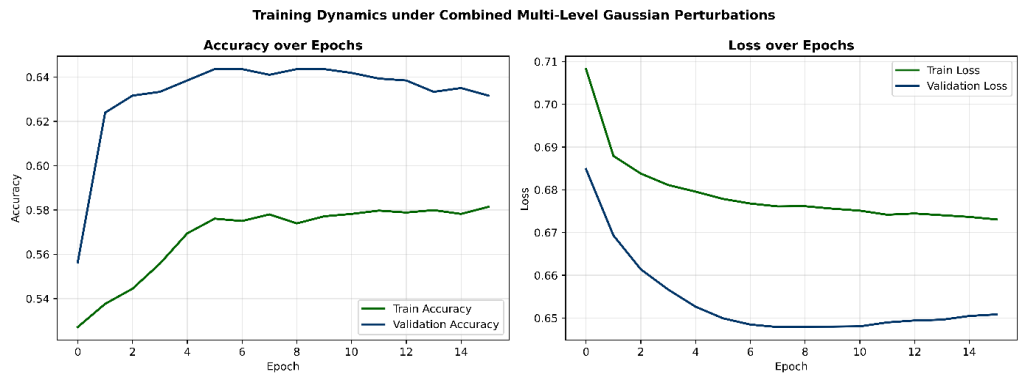

**Figure 26.** Epoch-wise variation of slice-level training and validation accuracy and loss for the specialized IA-QCNN architecture in predicting MGMT promoter methylation status under the designed hybrid noise scenario (multi-level Gaussian perturbations), where noise was applied simultaneously to both T1Gd MRI images and rotational gate blocks within the quantum circuit.

Furthermore, when transitioning from slice-level to patient-level analysis, the overall patient-level accuracy of the IA-QCNN model on the test set was obtained as 0.70 under the hybrid noise scenario. Under these conditions, the patient-level confusion matrix illustrating the MGMT promoter methylation prediction performance of the IA-QCNN architecture is presented in **Figure 27**, while the corresponding ROC curve is shown in **Figure 28**.

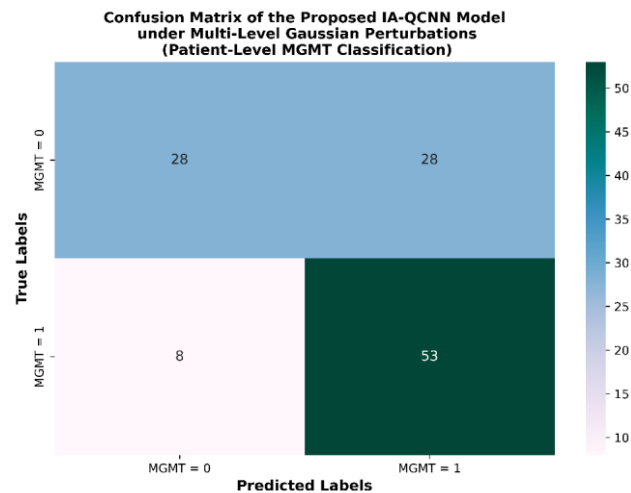

**Figure 27.** Patient-level confusion matrix of the specialized IA-QCNN architecture under the hybrid noise scenario (multi-level Gaussian perturbations) for MGMT promoter methylation status prediction.

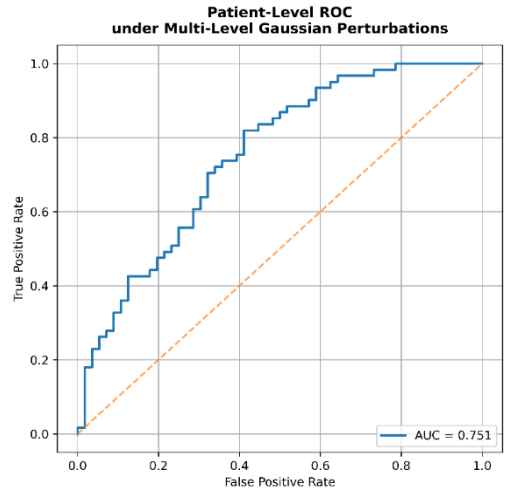

**Figure 28.** ROC curve and AUC value (AUC = 0.75) of the specialized IA-QCNN architecture under the hybrid noise scenario (multi-level Gaussian perturbations), obtained by aggregating slice-level predictions on a patient-level basis.

Under the hybrid noise scenario, the slice-level predictions of the specialized IA-QCNN architecture were aggregated into patient-level outcomes via averaging, yielding precision, recall, and F1-score values of 0.65, 0.87, and 0.75 for the methylated class, and 0.78, 0.50, and 0.61 for the unmethylated class, respectively. The detailed distribution of these class-wise metrics is illustrated in **Figure 29**.

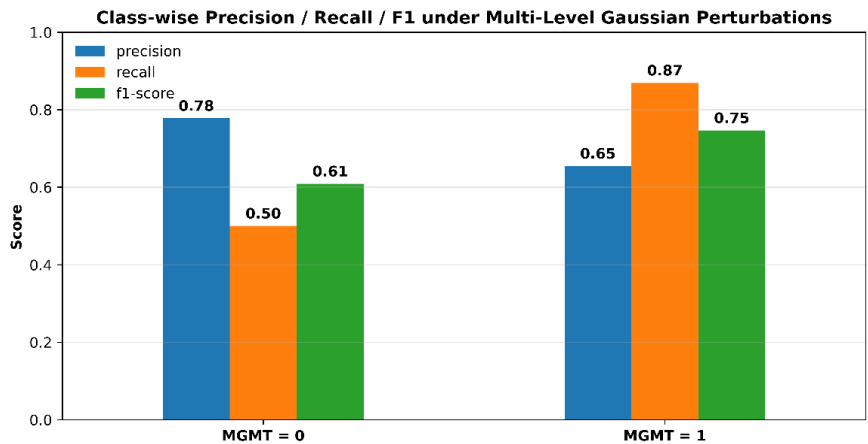

**Figure 29.** Detailed representation of class-wise precision, recall, and F1-score values for the specialized IA-QCNN architecture under the hybrid noise scenario (multi-level Gaussian perturbations).

### 3.6. Comparative results of CNN and DNN models designed with a similar number of parameters, along with Transfer Learning (TL) and State-of-the-Art (SOTA) models

In this stage of the study, the performance of the proposed specialized IA-QCNN architecture in predicting MGMT promoter methylation status—or its ability to distinguish between methylated and unmethylated classes—was compared against Convolutional Neural Network (CNN) and Deep Neural Network (DNN) architectures designed with similar parameter counts. As in the IA-QCNN architecture, the number of epochs, batch size, and learning rate for the classical CNN and DNN models were initially kept constant and set to 60, 16, and $3\times10^{-3}$, respectively. Subsequently, the ReduceLROnPlateau and EarlyStopping modules from the Keras library were incorporated into the training processes to enable dynamic optimization of the hyperparameters. The hidden layers of the DNN architecture consist of Dense layers with 64 and 32 neurons, respectively. A Dropout layer with a rate of 0.25 was applied between the first two layers to mitigate overfitting. The output layer of the model consists of two neurons corresponding to the methylated and unmethylated classes, with a softmax activation function. In addition, to ensure a fair comparison among the models, the input dimensionality of the classical CNN and DNN architectures was also determined in accordance with the qubit constraint of the quantum model (Equation 4). Thus, all three architectures were trained using the same number of features. When the DNN model was trained with the defined features, the training process was terminated at the 11th iteration via the Early Stopping mechanism. Under these conditions, the mean training accuracy, validation accuracy, training loss, and validation loss were recorded as 0.64, 0.58, 0.63, and 0.68, respectively; the corresponding accuracy and loss curves are illustrated in **Figure 30**.

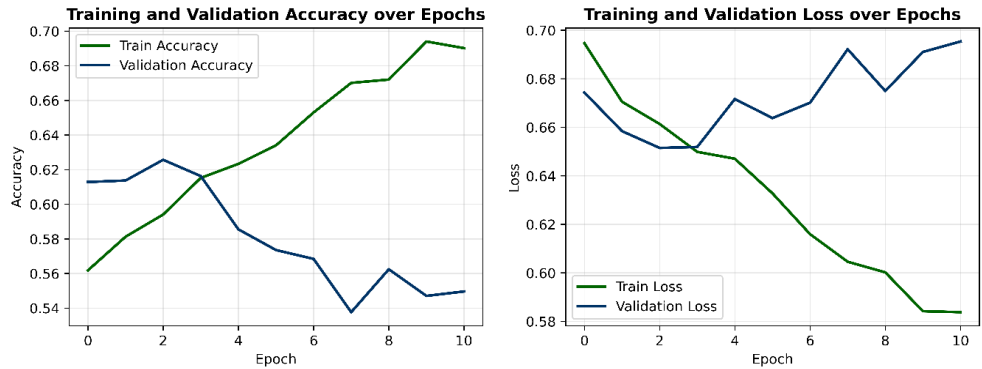

**Figure 30.** Epoch-wise variation of training and validation accuracy and loss for the DNN architecture designed with a similar number of parameters to the specialized IA-QCNN model for performance comparison.

In the subsequent stage, as in quantum-based approaches, slice-level prediction probabilities were aggregated into a patient-level representation in accordance with clinical decision-making requirements. As a result of this hierarchical evaluation, the overall patient-level accuracy of the DNN model on the test set was recorded as 0.64. The patient-level confusion matrix illustrating the performance of the DNN architecture in predicting MGMT promoter methylation status is presented in **Figure 31**, while the corresponding ROC curve is shown in **Figure 32**.

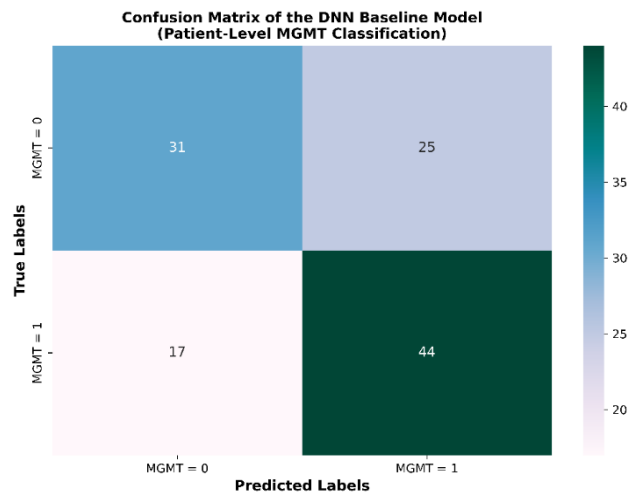

**Figure 31.** Patient-level confusion matrix of the DNN architecture for the prediction of MGMT promoter methylation status.

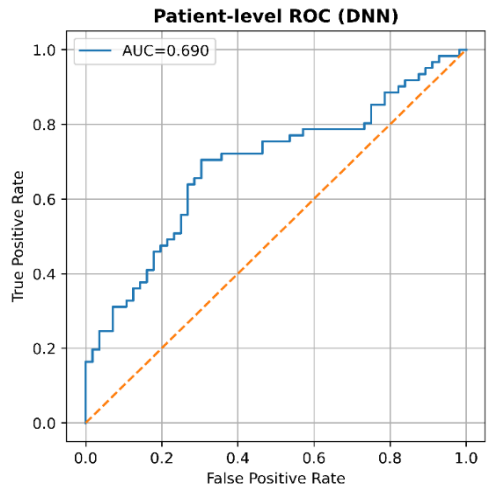

**Figure 32.** ROC curve and AUC value (AUC = 0.69) for the DNN architecture, obtained by aggregating slice-level predictions on a patient-level basis.

Furthermore, upon the patient-level mean aggregation of slice-level results obtained during the MGMT promoter methylation status prediction process of the DNN architecture, the precision, recall, and F1-score for the methylated class were recorded as 0.64, 0.72, and 0.68, respectively. For the unmethylated class, these values were found to be 0.64, 0.55, and 0.60, respectively (**Figure 36**).

The CNN model, trained under the same hyperparameter settings as the DNN architecture, consists of four Conv2D layers with 16, 32, 64, and 64 filters, respectively. In the convolutional layers, filters with a kernel size of 3×3 were employed, and the ReLU activation function was used. A 2×2 MaxPooling layer was applied after the first two convolutional layers, while a GlobalAveragePooling2D layer was introduced after the third convolutional layer to reduce the feature dimensionality. The final convolutional layer is followed by a Dropout layer with a rate of 0.25 to mitigate overfitting. The output layer consists of a Dense layer with two neurons, corresponding to the methylated and unmethylated classes, and uses a softmax activation function to produce class probabilities. The training process of the CNN model was terminated at the 35th epoch via the Early Stopping mechanism. Under these conditions, the mean training accuracy, validation accuracy, training loss, and validation loss were obtained as 0.56, 0.57, 0.68, and 0.68, respectively. The epoch-wise variation of accuracy and loss during the training and validation phases is presented in **Figure 33**.

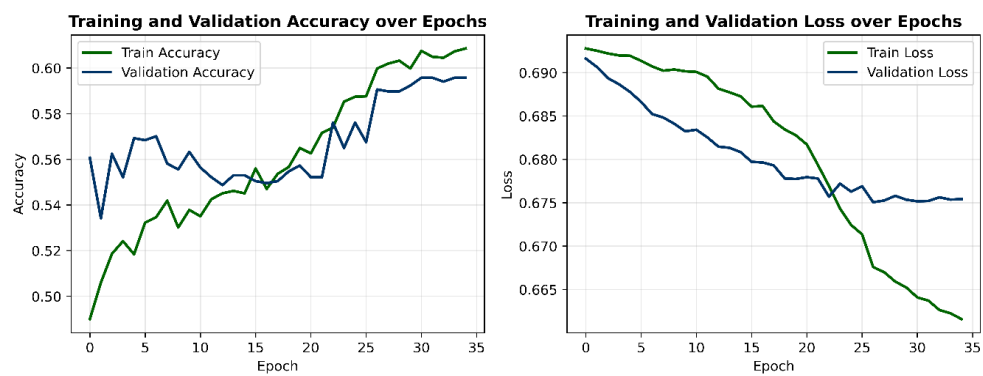

**Figure 33.** Epoch-wise variation of training and validation accuracy and loss for the CNN architecture designed with a similar number of parameters to the specialized IA-QCNN model for performance comparison.

Furthermore, upon transitioning from slice-level to patient-level results, the overall patient-level accuracy of the CNN architecture on the test set was recorded as 0.61. The patient-level confusion matrix, illustrating the performance of the CNN architecture in predicting MGMT promoter methylation status under these conditions, is presented in **Figure 34**, while the corresponding ROC curve is shown in **Figure 35**.

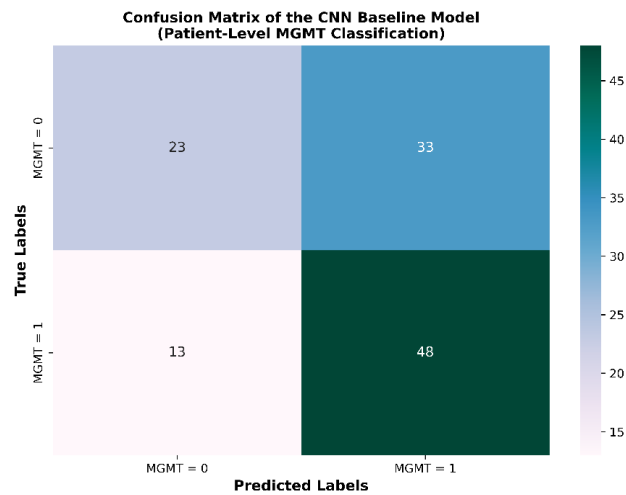

**Figure 34.** Patient-level confusion matrix of the CNN architecture for MGMT promoter methylation prediction.

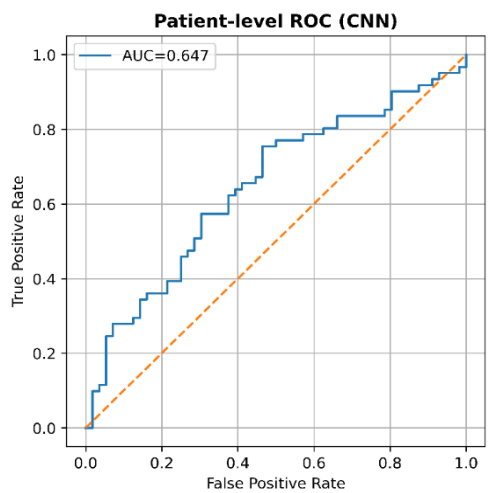

**Figure 35.** ROC curve and corresponding AUC value (AUC = 0.65) obtained by aggregating slice-level predictions of the CNN architecture into a patient-level representation.

In the patient-level evaluation of the CNN architecture, the precision, recall, and F1-score for the methylated class were recorded as 0.59, 0.79, and 0.68, respectively; for the unmethylated class, these values were determined to be 0.64, 0.41, and 0.50, respectively (**Figure 36**).

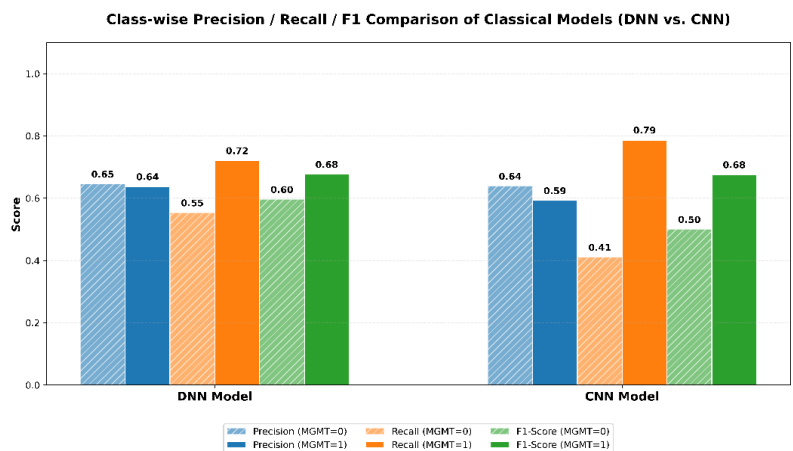

**Figure 36**. Detailed analysis of class-wise performance metrics (precision, recall, and F1-score) of the CNN and DNN models, designed for comparative performance evaluation against the specialized IA-QCNN architecture in predicting MGMT promoter methylation status.

In addition to the DNN and CNN models, the MGMT promoter methylation status prediction performance of the specialized IA-QCNN architecture was benchmarked against five robust classical TL models: DenseNet121, EfficientNetB0, ResNet50, InceptionV3, and VGG16. The performance of these models was evaluated using average training loss, validation loss, training accuracy, and validation accuracy metrics. The epoch-wise variation plots for each model are presented in **Supplementary Figure S6**, while the patient-level confusion matrices illustrating classification performance are provided in **Supplementary Figure S8**. Over 15 epochs, the average training and validation metrics were observed as follows: for DenseNet121, training accuracy of 0.73, training loss of 0.53, validation accuracy of 0.55, and validation loss of 0.74; for EfficientNetB0, training accuracy of 0.49, training loss of 0.69, validation accuracy of 0.52, and validation loss of 0.69; for InceptionV3, training accuracy of 0.78, training loss of 0.46, validation accuracy of 0.55, and validation loss of 0.85; for ResNet50, training accuracy of 0.51, training loss of 0.69, validation accuracy of 0.52, and validation loss of 0.69; and for VGG16, training accuracy of 0.61, training loss of 0.66, validation accuracy of 0.58, and validation loss of 0.68. The overall patient-level accuracies on the test set were calculated as 0.54 for DenseNet121, 0.52 for EfficientNetB0, 0.56 for InceptionV3, 0.51 for ResNet50, and 0.60 for VGG16. Class-wise precision, recall, and F1-score values are presented in **Supplementary Table S1**.

To further extend the scope of the study, the performance of the specialized IA-QCNN architecture was also compared with ConvNeXtTiny and EfficientNetV2S, which are among the contemporary high-performance SOTA DL architectures. As in the TL-based approaches, the performance of these models was evaluated using average training loss, validation loss, training accuracy, and validation accuracy metrics. The epoch-wise variation of accuracy and loss values during training and validation is presented in **Supplementary Figure S7**, while the corresponding patient-level confusion matrices are shown in **Supplementary Figure S9**. For ConvNeXt-Tiny, the training accuracy, training loss, validation accuracy, and validation loss were observed as 0.50, 0.69, 0.52, and 0.69, respectively; for EfficientNetV2-S, these values were 0.53, 0.69, 0.55, and 0.69, respectively. The overall patient-level test accuracies were determined as 0.52 for ConvNeXt-Tiny and 0.55 for EfficientNetV2-S. Class-wise precision, recall, and F1-score values are reported in **Supplementary Table S2**.

4. Discussion

Glioblastoma (GBM) tumors, characterized by pronounced intratumoral heterogeneity, aggressive biological behavior, and intense vascularization, are associated with limited therapeutic options, poor prognosis, and low survival rates, despite recent technological advances and an improved understanding of their pathophysiology. Determining the MGMT promoter methylation status, a predictive biomarker that significantly influences temozolomide-based treatment response and clinical outcomes in GBM patients, presents substantial diagnostic challenges for conventional analytical methods due to the complex and heterogeneous nature of the tumor. Currently, AI-based decision support systems are widely employed to overcome these challenges and to enable efficient non-invasive determination of MGMT promoter methylation status in GBM tumors using MRI-based neuroimaging data, achieving a certain level of success. Recent literature demonstrates a spectrum of methodologies with varying degrees of success. For instance, in this study [44], a multi-view approach was proposed to non-invasively determine MGMT promoter methylation using T2wi MRI scans from the RSNA-ASNR-MICCAI BraTS 2021 dataset, and the performance of the proposed model was compared with state-of-the-art methods. In the proposed model, a novel slice selection strategy was introduced to reduce the 3D MRI volume classification problem into a 2D classification problem. Following the preprocessing stage, segmentation was performed, and for each plane—axial, sagittal, and coronal—the slice containing the largest tumor region was selected based on the Feret diameter. The extracted slices were then subjected to feature extraction using MONAI DenseNet-121. The feature vectors obtained from the three planes for each patient were concatenated and fed into a three-layer fully connected classifier for MGMT methylation prediction. Consequently, the proposed model demonstrated a significant improvement in AUROC performance compared to state-of-the-art models and single-view approaches. In another study [91], a 3D Convolutional Neural Network (3D CNN)-based approach was proposed to predict the presence of MGMT promoter methylation using MRI images, and the performance of the proposed model was evaluated on the dataset provided in the RSNA-MICCAI Brain Tumor Radiogenomic Classification challenge. In the proposed model, an ensemble strategy was adopted in which four separate networks were constructed for each MRI sequence (FLAIR, T1w, T1wCE, T2), and their outputs were combined. Architecturally, a 3D ResNet-34 was employed. During training, 5-fold cross-validation was applied separately for each MRI sequence. The results demonstrated that the ensemble approach outperformed models trained on a single MRI modality, achieving AUC-ROC scores of 0.65010 on the public test set and 0.58295 on the private test set. In the study [47], in which MGMT promoter methylation status was predicted using T2 MRI images, three different approaches—voxel-wise, slice-wise, and whole-brain—were developed. The performance of each approach was evaluated using stratified 5-fold cross-validation according to MGMT status. In the voxel-wise approach, T2-weighted images were divided into $32\times32\times32$ patches, and a 3D V-Net model was employed for segmentation-based classification; the final decision was obtained via majority voting of voxel predictions within the largest connected component. In the slice-wise approach, a YOLOv5x model was used for object detection, whereas in the whole-brain approach, a 3D DenseNet-121 architecture was directly applied to the entire brain volume. The results showed that the voxel-wise approach achieved the highest performance, with an average accuracy of 65.42% (±3.97%) compared to the other methods. In another significant study [74], a bidirectional convolutional recurrent neural network (CRNN) architecture was proposed to predict MGMT promoter methylation status using axial brain MRI images of GBM patients, and the results were compared at both patient-level and MRI-level analyses. The MRI data used in the study were obtained from The Cancer Imaging Archive (TCIA), while the corresponding methylation data were sourced from The Cancer Genome Atlas (TCGA). In the proposed CRNN model, each MRI slice was individually fed into CNN layers, and the resulting feature vectors were subsequently processed as a sequence by a bidirectional RNN. Thus, the CRNN model was capable of handling MRI scans with a variable number of slices. Ultimately, when compared with a traditional Random Forest algorithm, which achieved an AUC of 0.44 on the test set, the proposed model demonstrated substantially superior performance, achieving 62% accuracy and 67% sensitivity.

According to the comprehensive literature review conducted within the scope of this study, no existing studies were identified that propose or develop quantum computing or quantum computing-based AI approaches for predicting MGMT promoter methylation status directly from MRI images in GBM tumors. Therefore, to address the imbalance between limited data availability in the existing literature and the high parameter burden of classical models, and to provide a novel perspective in GBM radiogenomic analysis, this study proposes a specialized framework exploiting the advantages of quantum computing. The performance of the proposed architecture in predicting MGMT promoter methylation status (i.e., distinguishing between methylated and unmethylated classes) based on MRI images was evaluated against both classical models reported in the literature and classical architectures trained under identical conditions (number of epochs, batch size, and learning rate). DNN and CNN architectures were designed to have parameter counts similar to the proposed quantum model. In addition, five robust transfer learning (TL) models—DenseNet121, EfficientNetB0, ResNet50, InceptionV3, and VGG16—obtained from the Keras library were evaluated in a comparative framework. Alongside these TL methods, ConvNeXtTiny and EfficientNetV2S, which are among the most recent high-performance deep learning architectures, were also incorporated into the MGMT promoter methylation prediction pipeline for GBM tumors, and the performance of all classical models was benchmarked against the proposed architecture.

In this study, a problem-specific quantum deep learning architecture has been proposed in order to establish a methodological bridge between GBM radiogenomic analysis and quantum deep learning. To the best of our knowledge, this study presents the first comprehensive quantum-based approach developed for predicting MGMT promoter methylation status directly from MR images in GBM tumors, thereby addressing a significant gap in the existing literature. The primary methodological contribution of this work is the specialized Importance-Aware Quantum Convolutional Neural Network with Ring-Topology (specialized IA-QCNN) architecture, which can be directly applied to the analysis of highly heterogeneous and complex tumors such as GBM and can determine MGMT promoter methylation status from MR images, overcoming the limitations of classical deep learning models. Unlike existing approaches in the literature, the proposed model is distinguished by three fundamental methodological innovations. First, in the preprocessing stage, an energy-based slice selection method is proposed in order to identify tumor-representative slices without segmentation and to enable the systematic and reproducible selection of the most clinically informative slices containing radiogenomic information. Using this energy-based slice selection method, T1Gd MRI slices for each patient are systematically ranked according to their computed energy scores. Second, within the importance-aware weighting component of the specialized IA-QCNN architecture, instead of the position-based weighting strategy used in standard CNN and QCNN models, an importance-aware weighting approach is adopted that adaptively scales the contribution of features derived from pixel intensity information to the quantum representation obtained via angle-based feature encoding and unitary transformations. In this way, while the model determines which features are more or less important for predicting MGMT promoter methylation status, critical local patterns reflecting the biological heterogeneity of the tumor are automatically emphasized throughout the training process. Another key contribution of the study is its novel architecture, which takes into account the constraints of current NISQ devices and integrates ring-topology quantum convolution and folding-based quantum pooling layers. In the ring-topology quantum convolution layer, the weight-sharing principle of classical CNNs is transferred into the quantum domain, and a global correlation is established among neighboring qubits through a ring topology, while simultaneously optimizing the number of trainable parameters. The folding-based quantum pooling layer, based on unitary transformations, minimizes information loss during hierarchical feature extraction, thereby improving both the robustness to noise and the computational efficiency of the model.

The total number of trainable parameters obtained when the proposed specialized IA-QCNN architecture is executed using TensorFlow Quantum (TFQ), which performs state-vector simulations on the Cirq framework, is given as $N_{qconv} + N_{qpool} + N_{importance\_aware\_weighting} + N_{dense} = 6 + 3 + 2\ x\ 18 + 10 = 55$. The quantum convolution ($N_{qconv}$) and pooling ($N_{qpool}$) layers contribute 6 and 3 parameters, respectively, due to the weight-sharing principle. The importance-aware weighting mechanism introduces $2d$ parameters via $R_y$ and $R_z$ rotational gates, where $d$ denotes the number of qubits automatically determined after PCA and is constrained to a maximum of $d_{MAX} = 18$. Here, the total number of parameters is computed using the maximum value. The remaining 10 parameters in the dense layer ($N_{dense}$) are calculated using the $(input \times output) + bias$ formulation. This compact parameterization enhances learning efficiency while preserving the expressive capacity of the model. The prediction of MGMT promoter methylation status in GBM tumors using the specialized IA-QCNN was initially conducted on T1Gd MRI images. At the end of the 17th iteration, the average training and validation accuracies were 0.56 and 0.62, respectively, while the training and validation losses were 0.69 and 0.67. Furthermore, when slice-based prediction probabilities were aggregated at the patient level, the overall patient-level accuracy of the specialized IA-QCNN on the test set reached 0.67. Among the classical models, the DNN and VGG16 architectures achieved the highest validation accuracy (0.58) and the lowest validation loss (0.68) compared to other classical benchmarks. However, the DNN architecture exhibited superior test performance with a patient-based accuracy of 0.64 compared to VGG16. Despite these results, classical models generally exhibited relatively low validation accuracy and high validation loss, indicating significant overfitting and limited generalization capabilities. In contrast, the more balanced distribution between training and validation performance in the proposed specialized IA-QCNN demonstrates superior generalization. Moreover, in terms of trainable parameters, the specialized IA-QCNN utilizes only 55 parameters—representing a 98% and 99% reduction compared to DNN (2,978) and VGG16 (65,922), respectively. Despite this drastic reduction, it achieved a 4.7% higher patient-level accuracy than DNN and an 11.7% higher accuracy than VGG16 on the test set. The fact that the quantum models outperform classical models with significantly fewer parameters aligns with several previous studies in the literature [57,58,92,93]. These results suggest that the specialized IA-QCNN not only minimizes computational resources but also provides a robust solution to the overfitting problems encountered by even SOTA DL architectures in high-dimensional, complex GBM radiogenomic data. This indicates that the model transcends mere data memorization and effectively assimilates the fundamental radiogenomic patterns specific to MGMT methylation through its low-parameter yet high-representational capacity in quantum space.

One of the key points to be emphasized here is that the IA-QCNN architecture achieves higher prediction performance with significantly fewer parameters compared to classical models, which is not only a result of computational efficiency but also a consequence of both the methodological innovation introduced by the proposed model and the high data representation capability provided by the quantum Hilbert space. In the importance-aware weighting approach of the proposed specialized IA-QCNN architecture, features derived from T1Gd pixel intensity information are first encoded as quantum states via angle-based feature encoding (through the $R_y$ rotational gate), after which a mechanism is developed to scale the contribution of these data to the quantum representation using end-to-end learnable and differentiable parameters. One of the primary motivations for developing this problem-specific approach lies in the spatially variable and heterogeneous nature of MGMT status in GBM tumors, which necessitates shifting the research perspective away from the "where the tumor is located" paradigm toward focusing on the "biological meaning of the underlying tissue signal". In the proposed approach, features derived from T1Gd pixel intensities are encoded into the amplitude and phase components of the quantum state via $R_y$ and $R_z$ rotational gates, and these contributions are dynamically adjusted through a learnable scaling mechanism dependent on angular parameters. In this way, the model moves beyond the classical position-oriented approaches and dynamically determines "what to focus on" by emphasizing the most discriminative radiogenomic biomarkers directly associated with MGMT methylation status throughout the training process (in other words, by suppressing noise). Another methodological innovation introduced by the proposed specialized IA-QCNN architecture is the ring-topology quantum convolution and folding-based quantum pooling layers. The quantum convolution layer consists of rotational blocks defined with shared parameters between neighboring qubit pairs and CNOT gates used to establish entanglement. Here, similar to the classical weight-sharing principle, the number of trainable parameters is significantly optimized through the design of the quantum convolution block; additionally, the ring topology enables the establishment of global correlations among all qubits, allowing critical radiogenomic information in GBM tissues to be effectively propagated throughout the circuit with minimal parameter overhead. In the quantum pooling layer, similarly to the quantum convolution layer, parameters are shared across rotational blocks between neighboring qubit pairs, thereby minimizing the number of trainable parameters. Thus, while the number of qubits is reduced through a "folding-based quantum pooling" strategy, the loss of radiogenomic information critical for MGMT methylation prediction is prevented, and the robustness of the model against noise is enhanced. In addition to the methodological innovations introduced by the IA-QCNN architecture, the large information capacity and high data representation capability provided by the Hilbert space also form the foundation of the achieved parameter efficiency and classification performance.

Quantum computers possess the potential to analyze highly correlated and complex data relationships—which are difficult to resolve on classical computers—more efficiently and accurately within the high-dimensional Hilbert space [94]. In the case of heterogeneous and highly correlated data, such as GBM tumors, both the data itself and the relationships among data points are represented within a more limited space in classical vector representations. In contrast, the expansive representational capacity of high-dimensional quantum Hilbert space enables more effective capture of microscopic tissue variations that may be overlooked by classical models, thereby allowing a more precise determination of the spatially distributed MGMT status in GBM tumors. When the performance of the proposed specialized IA-QCNN architecture in predicting MGMT methylation status in GBM tumors based on T1Gd MRI images is illustrated using representative samples (**Supplementary Figure S10**), it can be observed that, in the TP case, the model assigns a high prediction probability ($p = 0.751$), successfully capturing critical radiogenomic features within the tumor slice. The low probability value assigned to the TN case ($p = 0.346$) further demonstrates the model's ability to effectively distinguish between methylated and unmethylated states. The errors observed in the FP and FN cases are likely attributable to the heterogeneous nature of GBM tumors or to the misrepresentation of features that are not biologically significant at the microscopic level, which may be incorrectly emphasized in the Hilbert space through the importance-aware weighting mechanism. An examination of the training and validation accuracy/loss curves reveals that the gap between training and validation accuracy, along with the increase in validation loss observed after approximately the 10th epoch, indicates a mild degree of overfitting emerging in the later stages of training. This phenomenon is largely driven by data limitations, which restrict the model's ability to generalize across diverse patterns. Furthermore, despite the relatively complex structure of the proposed specialized IA-QCNN architecture, the limited size of the dataset used during training and validation is also considered a significant factor constraining the model's generalization capability.

In this study, the performance of the proposed specialized IA-QCNN architecture in predicting MGMT methylation status in GBM tumors was also evaluated using multi-parametric (mp) MRI modalities, including FLAIR, T1w, and T2, and the prediction results obtained from mpMRI and T1Gd MRI were compared. The training process conducted using mpMRI data was terminated at the 12th iteration via the Early Stopping mechanism, yielding average training and validation accuracies of 0.59 and 0.46, respectively, and training and validation losses of 0.67 and 0.74, respectively. Compared with the T1Gd MRI modality, the use of mpMRI led to an approximately 5.4% increase in training accuracy, whereas validation accuracy decreased by about 25.8%. Meanwhile, training and validation losses exhibited a decrease of 2.9% and an increase of 10.4%, respectively. Furthermore, an important point to emphasize here is that, when examining the training and validation accuracy/loss curves obtained using mpMRI, a noticeable gap between training and validation performance is observed; moreover, the near-constant validation accuracy and the increase in validation loss after a certain number of iterations indicate a substantial decline in generalization capability and a high risk of overfitting. These findings suggest that combining multiple MRI sequences negatively affects the model's learning capability in determining MGMT methylation status in GBM tumors. Considering the spatially heterogeneous nature of MGMT status in GBM tumors, the contrast-enhanced, active, vascularized, and blood–brain barrier–disrupted regions highlighted in T1Gd MRI may play a more effective role in determining MGMT status, representing a clinically meaningful sequence preference supported by the findings of this study. In conclusion, while this sequence preference does not exclude the widely reported success of mpMRI-based approaches in the literature, it demonstrates that the proposed specialized IA-QCNN architecture yields more meaningful and stable results with the T1Gd modality for determining the radiogenomic profile of GBM tumors.

In the final stage of this study, the proposed specialized IA-QCNN architecture was evaluated under three different noise environments in order to assess its suitability for real-world conditions, its reliability, and to analyze its behavior in settings where gate-induced errors inherent to real quantum computing environments are approximately simulated. In the first scenario, Gaussian-distributed random noise with zero mean and a standard deviation of 0.03 was added to the pixel intensities of all MRI images, including T1Gd, and the model was tested under a mild-to-moderate noise environment. In the second scenario, Gaussian-distributed random noise with a standard deviation of 0.02 was injected into the rotation angles of all quantum gates throughout the angle-based quantum feature encoding, importance-aware weighting, quantum convolution, and pooling layers of the IA-QCNN architecture. In the final scenario, a hybrid noise environment was designed, and the model was evaluated under mild-to-moderate noise simultaneously applied to both pixel intensities and quantum gate parameters. Transitioning from the first scenario to the hybrid noise scenario, an increase of 1.79% in training accuracy and 10.34% in validation accuracy was observed. While the training loss remained unchanged, the validation loss decreased by approximately 2.99%. In addition, a 14.75% improvement in patient-level prediction accuracy was achieved. These results demonstrate that the proposed model remains robust even under conditions where both characteristic gate errors of current NISQ quantum computing environments and data perturbations encountered in real clinical settings are jointly simulated, indicating a high degree of "noise immunity". Another notable finding of this study is that, under the hybrid noise scenario, the model's AUC performance increased to 0.75. In this setting, the IA-QCNN architecture not only develops robustness against noise but also effectively utilizes it as a "stochastic regularization mechanism", analogous to the dropout technique in classical ANN architectures, thereby avoiding local minima. This behavior facilitates the distinction between methylated and unmethylated states while simultaneously reducing the risk of overfitting observed in other scenarios. On the other hand, despite the observed improvement in AUC performance, the noticeable gap between training and validation curves persists due to the limited size of the dataset.

In conclusion, the proposed specialized IA-QCNN architecture demonstrates high performance and strong generalization capability in predicting MGMT promoter methylation status in GBM tumors based on T1Gd MRI images, despite having a significantly lower number of trainable parameters compared to classical DL models, and provides a robust solution to the overfitting problem encountered by classical SOTA architectures. Moreover, when compared to mpMRI, the T1Gd MRI modality was found to play a more effective role in determining MGMT status in GBM tumors and is therefore proposed as a clinically meaningful sequence preference within the scope of this study. Notably, under the hybrid noise environment, the proposed model not only develops robustness to noise but also utilizes it as a regularization mechanism, thereby improving performance. Overall, the findings demonstrate that the proposed specialized IA-QCNN architecture constitutes a strong and reliable alternative to classical approaches in the analysis of complex and heterogeneous radiogenomic data, and reveals the quantum advantage not only at a theoretical level but also in practical applications.

## 5. Conclusions, Limitations and Future Works

In this study, a specialized Importance-Aware Quantum Convolutional Neural Network with Ring-Topology (IA-QCNN) architecture was proposed to predict MGMT promoter methylation status in GBM tumors directly from MRI images, and its performance was evaluated on the RSNA-MICCAI brain tumor radiogenomic dataset. The proposed model represents the first comprehensive quantum-based approach in this domain by establishing a methodological connection between GBM radiogenomic analysis and quantum deep learning. The specialized IA-QCNN architecture was evaluated using both mpMRI and T1Gd MRI images for MGMT promoter methylation prediction and, by more effectively representing contrast-enhanced active regions, contributes to the literature not only methodologically but also through a clinically meaningful sequence preference. Furthermore, when the proposed model was tested under mild-to-moderate Gaussian noise simulating both characteristic gate errors in NISQ computing environments and data perturbations encountered in real clinical settings, its performance showed a moderate decline in both cases; however, under the hybrid noise scenario, the model utilized noise as a regularization mechanism, avoided local minima, and achieved improved performance in determining MGMT promoter methylation compared to the baseline condition. Finally, when compared with classical DNN/CNN with similar parameter counts, transfer learning models, state-of-the-art DL architectures, and related studies in the literature, the proposed IA-QCNN model demonstrates that it constitutes a viable alternative to classical approaches for the analysis of complex radiogenomic data, owing to its high performance and strong generalization capability achieved with substantially fewer parameters.

In the conclusion section of the manuscript, it is appropriate to state the following potential limitations of the study:

**Dataset Limitation:** In this study, the prediction of MGMT methylation status in GBM tumors was performed using the RSNA-MICCAI Brain Tumor Radiogenomic dataset, which was released as part of the Brain Tumor Radiogenomic Classification global challenge. Since no other publicly available dataset comparable to this dataset exists in the literature, the study was conducted solely on this dataset. However, within the dataset, analyses were performed using different variations (T1Gd, mpMRI, and noise models), and the model performance was evaluated across these variations. Nevertheless, to fully validate the generalization capability of the proposed IA-QCNN architecture, it is necessary to test it on external heterogeneous datasets obtained from different populations and MRI devices.

**Fixed-Depth Quantum Circuit Usage:** In this study, a fixed-depth quantum circuit structure was intentionally used in the IA-QCNN in order to reduce computational time and model complexity. Since the effect of quantum pooling increases as the circuit depth grows, the circuit depth was kept constant to maintain a balance between the number of input qubits and the number of output features. However, using the pseudo-code of the proposed architecture provided in the supplementary materials, the quantum circuit depth can be increased if required.

**Cirq Noise Channels and Models:** Various built-in noise models within the Cirq simulation environment were initially implemented. However, due to the inherent complexity of the proposed model—particularly in circuits with high qubit counts and deep layers—these built-in models significantly increased simulation times and, in some instances, led to kernel failures; these constraints were observed despite the utilization of supercomputing infrastructure (Perlmutter-NERSC). Consequently, Gaussian noise was preferred for simulations, as this approach simplified noise modeling while substantially reducing processing time.

**Barren Plateaus:** In general, the barren plateau problem is often overlooked in QCNN architectures; however, the applied noise models and increasing number of qubits may affect gradient stability. Although this risk was mitigated in this study through a fixed-depth architecture and ring-topology design, preserving gradient dynamics in more complex and noisy hardware environments remains an open issue for future work.

In future studies, we aim to extend the specialized IA-QCNN into a fully end-to-end quantum model that operates without the need for classical dimensionality reduction techniques, such as PCA, during intermediate stages. Additionally, we intend to develop a novel quantum representation method to further enhance the model's analytical power.

**CRediT Authorship Contribution Statement**
**EA:** Conceptualization, Methodology, Software, Validation, Formal analysis, Investigation, Resources, Data curation, Writing – original draft, Writing – review & editing, Visualization, Funding acquisition. **MO:** Writing – review & editing, Funding acquisition, Supervision. **All authors:** Read and approved the final manuscript.

**Funding**
This study was supported by the Scientific and Technological Research Council of Turkey (TUBITAK) [grant number 124F213]. The authors thank TUBITAK for their support. This work was also supported by the Scientific and Technological Research Council of Turkey (TUBITAK) under the 2211/C National Ph.D. Scholarship Program in Priority Fields in Science and Technology.

**Data Availability Statement**
The dataset utilized in this study is the RSNA-MICCAI Brain Tumor Radiogenomic dataset, published on the Kaggle platform as part of the 'Brain Tumor Radiogenomic Classification' competition. This competition was organized through the collaboration of the Radiological Society of North America (RSNA) and the Medical Image Computing and Computer-Assisted Intervention (MICCAI) Society. The data is publicly accessible at: https://www.kaggle.com/competitions/rsna-miccai-brain-tumor-radiogenomic-classification

**Code Availability Statement**
The code supporting the findings of this study is not publicly available at the time of submission; however, it will be made publicly available via a GitHub repository upon publication of the article.

**Acknowledgments**
The authors would like to acknowledge that this paper is submitted in partial fulfillment of the requirements for PhD degree at Yildiz Technical University.

This research used resources of the National Energy Research Scientific Computing Center (NERSC), a DOE Office of Science User Facility supported by the Office of Science of the U.S. Department of Energy under Contract No. DE-AC02-05CH11231, using NERSC awards DDR-ERCAP0033396 and DDR-ERCAP0038319.

**Declaration of Competing Interests**

The authors declare that they have no known competing financial interests or personal relationships that could have appeared to influence the work reported in this paper.

# SUPPLEMENTARY MATERIAL

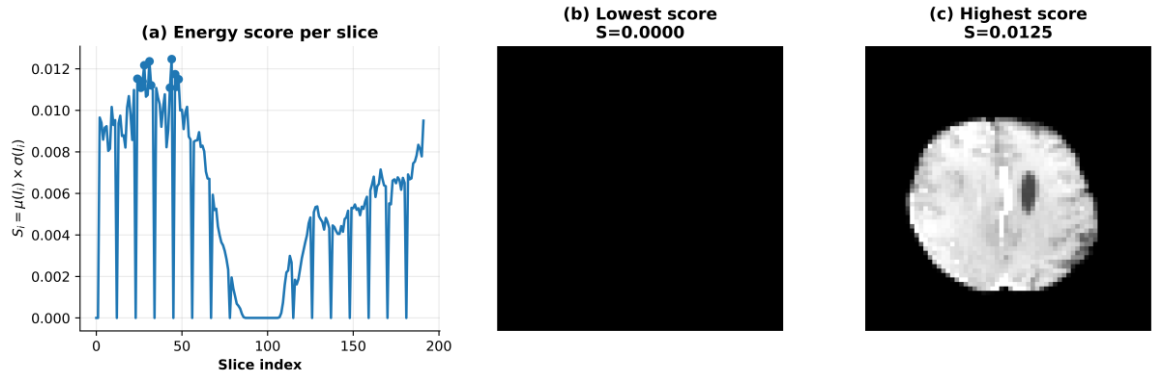

**Supplementary Figure S1.** Visualization of the proposed energy-based slice selection strategy for Case 00400 from the training dataset. **(a)** Distribution of energy scores for each slice, computed using the approach $S_i = \mu(I_i) \times \sigma(I_i)$, as a function of the slice index. Based on this distribution, the top K = 10 slices with the highest energy scores correspond to slices with indices 44, 31, 28, 46, 24, 48, 27, 32, 26 and 43. **(b)** The slice with the lowest energy score represents regions with minimal anatomical content, predominantly consisting of background information. **(c)** The slice with the highest energy score illustrates the slice selected by the proposed method, in which anatomical structures and clinically informative brain regions are prominently observable.

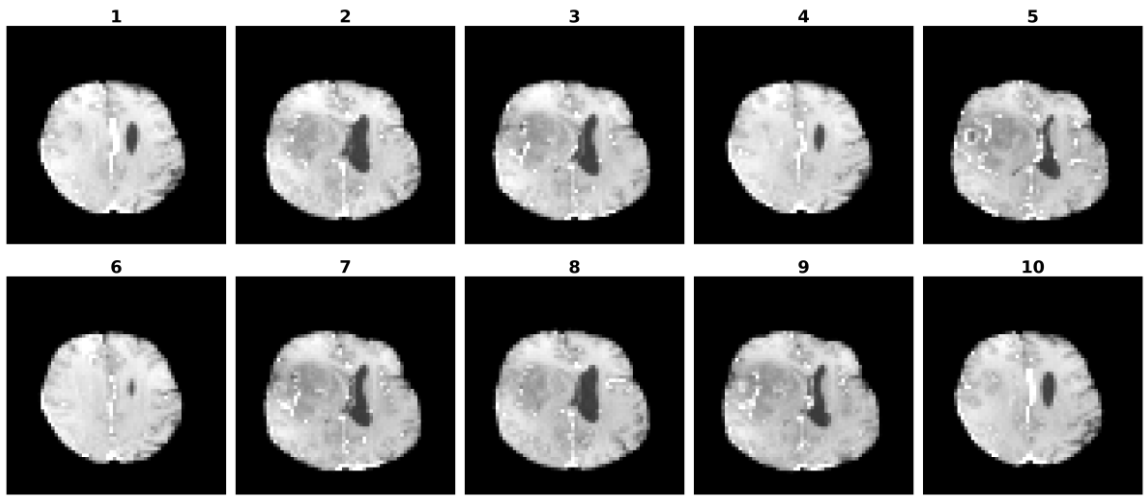

**Supplementary Figure S2.** The top K = 10 slices selected based on slice-wise energy scores computed using the proposed energy-based slice selection method, along with their corresponding T1Gd MRI representations.

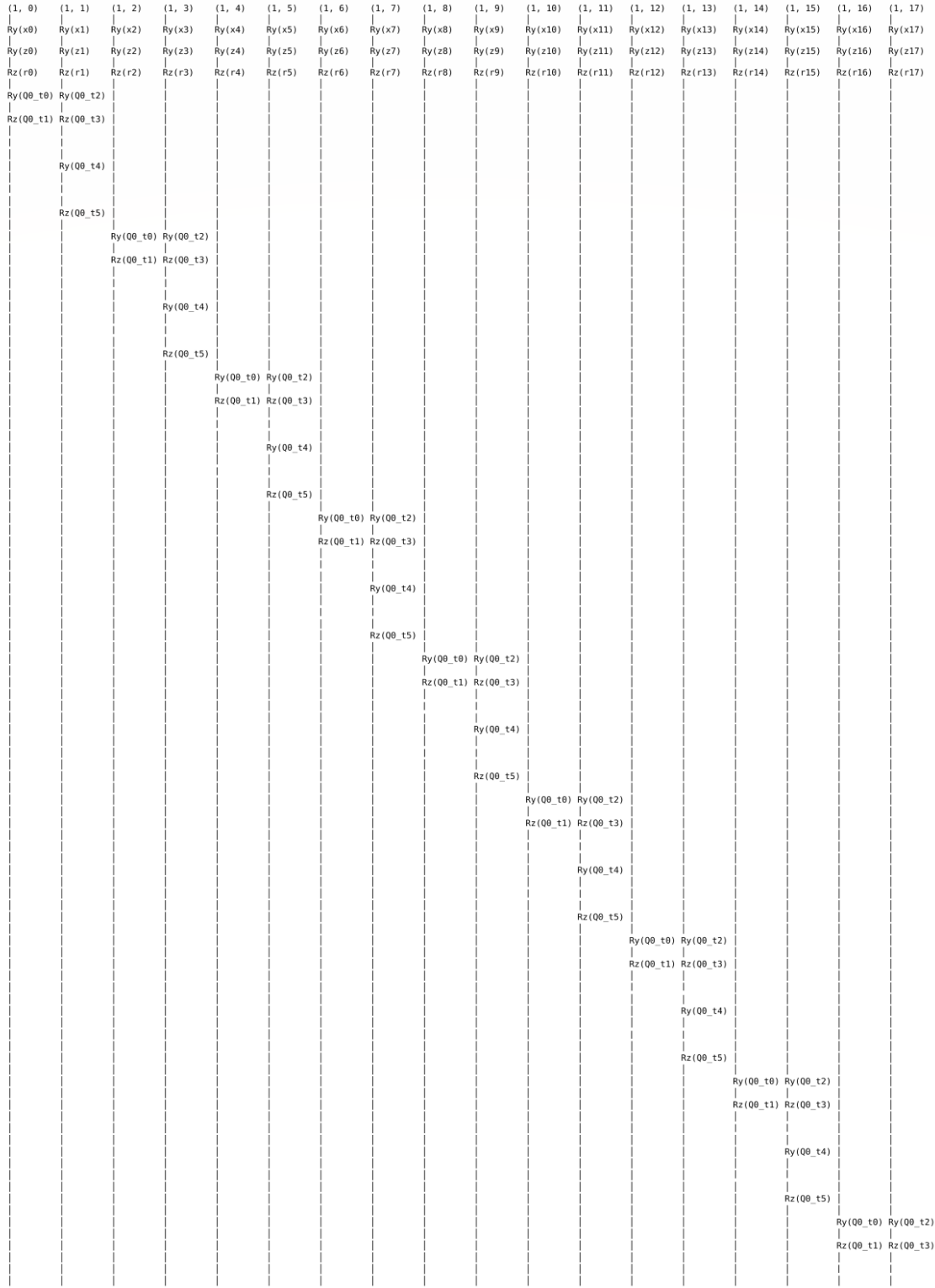

**Supplementary Figure S3.** Full schematic of the quantum circuits constructed using TensorFlow Quantum (TFQ) and the Cirq framework for the proposed angle-based quantum feature encoding and importance-aware weighting strategy. The number of qubits is automatically determined and constrained by the expression $d_{qubits} = min(d_{PCA}, d_{MAX})$.

## Algorithm 2: Ring-Topology Quantum Convolution and Folding-Based Quantum Pooling

```
Input:
    F = {q_1, q_2, ..., q_N}                    // active qubits at current level
    L                                            // number of QCNN levels

Output:
    F_active                                     // remaining pooled qubits
    QCNN                                         // QCNN circuit

Subroutine QCONV_BLOCK(q_a, q_b; t_0..t_5):
    Apply Ry(t_0) on q_a
    Apply Rz(t_1) on q_a
    Apply Ry(t_2) on q_b
    Apply Rz(t_3) on q_b
    Apply CNOT(q_a → q_b)
    Apply Ry(t_4) on q_b
    Apply CNOT(q_a → q_b)
    Apply Rz(t_5) on q_b
    return block

Subroutine RING_QCONV_LAYER(F; t_0..t_5):
    Initialize U as an empty layer

    for k = 1 to n step 2 do
        U ← U ⊕ QCONV_BLOCK(F[k], F[k+1]; t_0..t_5)
    end for

    for k = 2 to n-1 step 2 do
        U ← U ⊕ QCONV_BLOCK(F[k], F[k+1]; t_0..t_5)
    end for

    U ← U ⊕ QCONV_BLOCK(F[n], F[1]; t_0..t_5)
    return U

Subroutine POOL_BLOCK(source, target; p_0..p_2):
    Apply Ry(p_0) on target
    Apply Rz(p_1) on source
    Apply CNOT(source → target)
    Apply Ry(p_2) on target
    return block

Main QCNN_CORE(F, L):
Initialize QCNN as an empty circuit
 F_active ← F

    for lvl = 0 to L-1 do
        θ_lvl ← (t_0..t_5)
        φ_lvl ← (p_0..p_2)

        QCNN ← QCNN ⊕ RING_QCONV_LAYER(F_active; θ_lvl)   //Ring quantum

        n ← |F_active|                           //Pooling operation
        half ← floor(n / 2)

        source ← F_active[1 : half]
        target ← F_active[half+1 : n]

        for i = 1 to half do
            QCNN ← QCNN ⊕ POOL_BLOCK(source[i], target[i]; φ_lvl)
        end for

        F_active ← target                        //Downsampling

    end for

    return (QCNN, F_active)
```

**Supplementary Figure S4.** Pseudocode of the algorithm defining the proposed ring-topology quantum convolution and folding-based quantum pooling mechanisms.

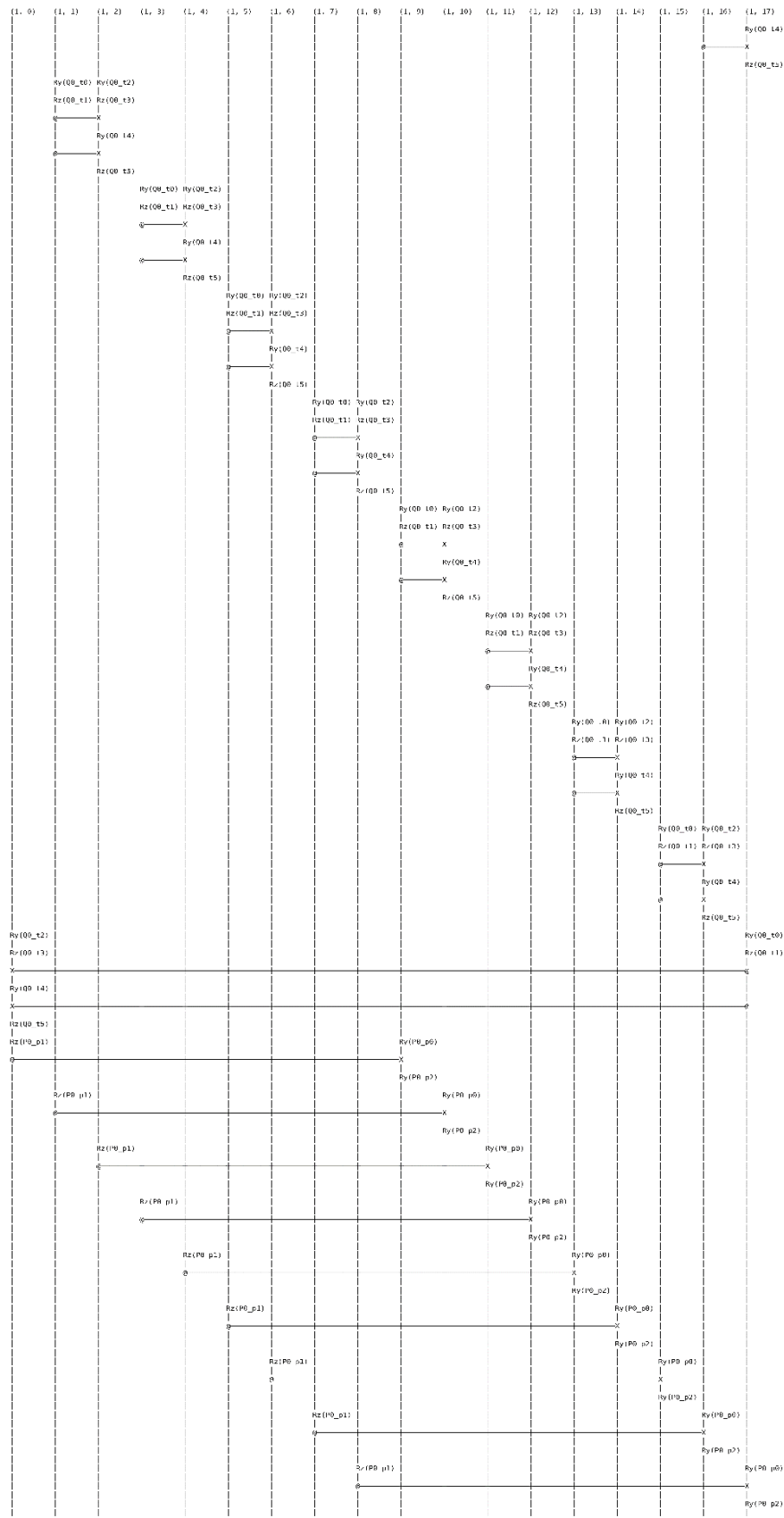

**Supplementary Figure S5.** Full schematic of the quantum circuits constructed using TensorFlow Quantum (TFQ) and the Cirq framework for the proposed ring-topology quantum convolution and folding-based quantum pooling approaches.

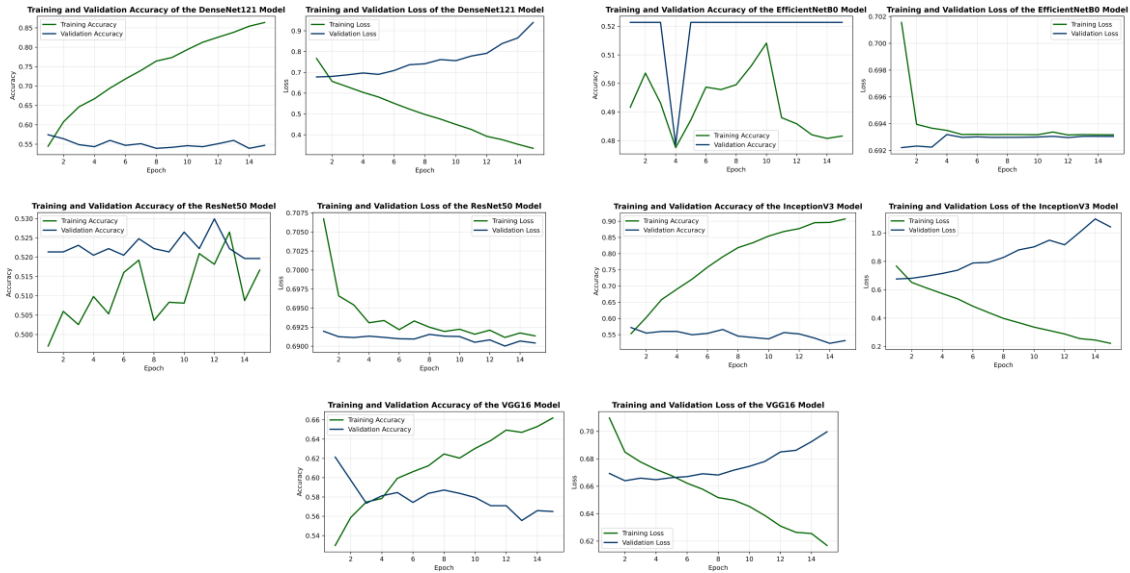

**Supplementary Figure S6.** Epoch-wise variation of training and validation accuracy and loss for the prediction of MGMT promoter methylation status in GBM tumors using five different classical transfer learning models—DenseNet121, EfficientNetB0, ResNet50, InceptionV3, and VGG16.

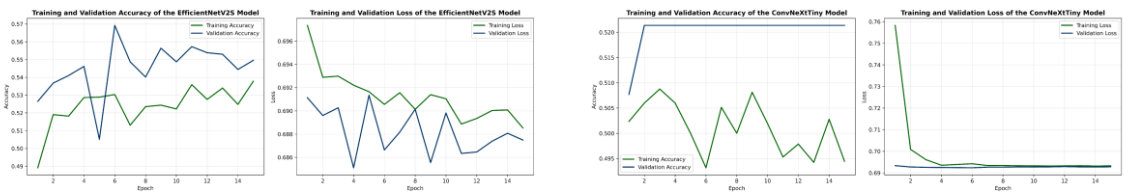

**Supplementary Figure S7.** Epoch-wise variation of training and validation accuracy and loss for the prediction of MGMT promoter methylation status in GBM tumors using ConvNeXtTiny and EfficientNetV2S models.

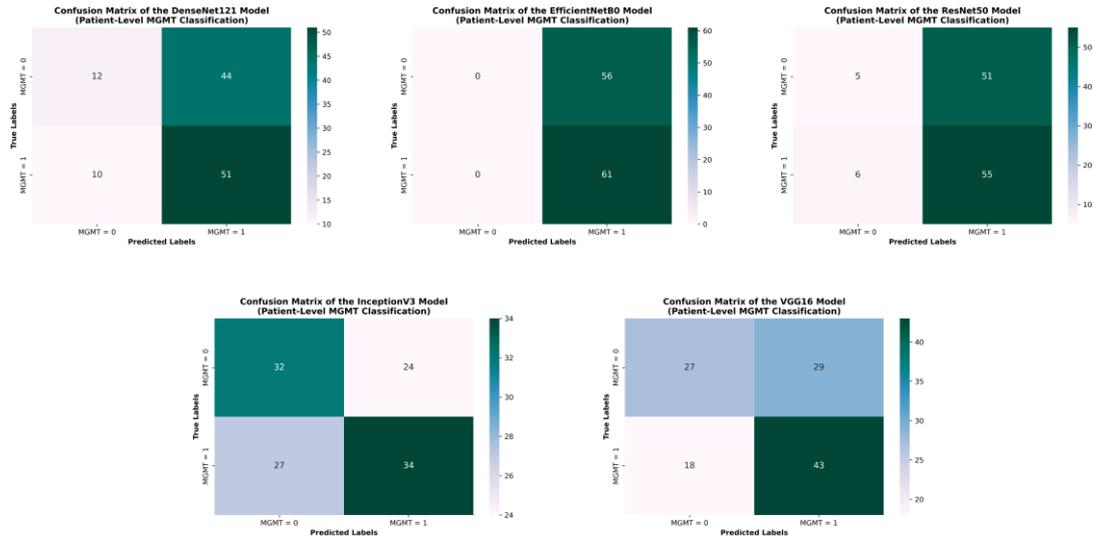

**Supplementary Figure S8.** Confusion matrices for the prediction of MGMT promoter methylation status in GBM tumors using DenseNet121, EfficientNetB0, ResNet50, InceptionV3, and VGG16 models.

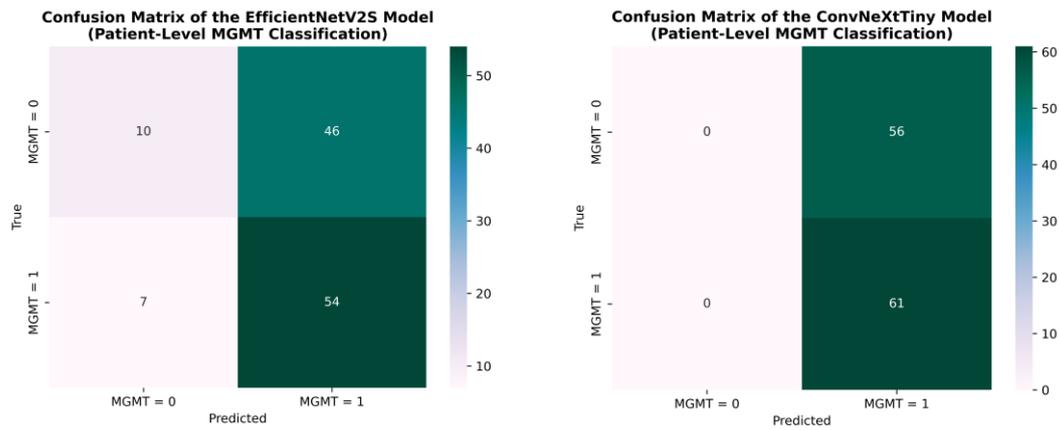

**Supplementary Figure S9.** Confusion matrices for the prediction of MGMT promoter methylation status in GBM tumors using EfficientNetV2S AND ConvNeXtTiny models.

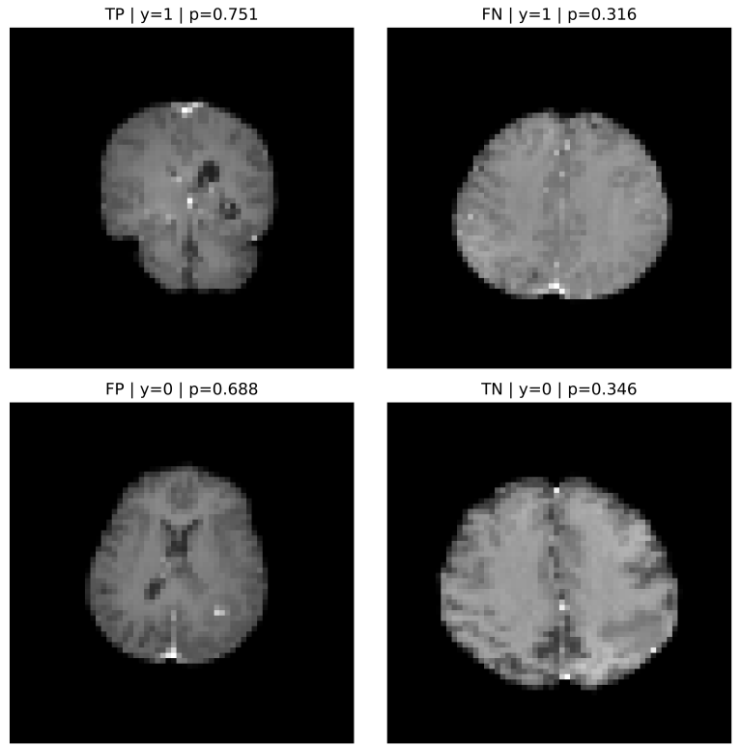

**Supplementary Figure S10.** Qualitative assessment of MGMT promoter methylation status predictions obtained from the proposed specialized IA-QCNN model on T1Gd MRI-based GBM tumor samples. Representative T1Gd MRI slices for True Positive (TP), False Positive (FP), False Negative (FN), and True Negative (TN) cases are presented along with the corresponding ground-truth labels (y) and predicted probabilities (p).

**Supplementary Table S1.** Detailed performance metrics of five distinct and robust classical transfer learning (TL) architectures (DenseNet121, EfficientNetB0, ResNet50, InceptionV3, and VGG16) for predicting MGMT promoter methylation status in GBM tumors. The table presents class-specific precision, recall, and F1-score values for each model.

|  |  | **Precision** | **Recall** | **F1-Score** |
|---|---|---|---|---|
| **DenseNet121** | *MGMT-Methylated* | 0.54 | 0.84 | 0.65 |
|  | *MGMT-Unmethylated* | 0.55 | 0.21 | 0.31 |
| **EfficientNetB0** | *MGMT-Methylated* | 0.52 | 1.0 | 0.69 |
|  | *MGMT-Unmethylated* | 0.1 | 0.1 | 0.1 |
| **ResNet50** | *MGMT-Methylated* | 0.52 | 0.90 | 0.66 |
|  | *MGMT-Unmethylated* | 0.45 | 0.09 | 0.2 |
| **InceptionV3** | *MGMT-Methylated* | 0.59 | 0.56 | 0.57 |
|  | *MGMT-Unmethylated* | 0.54 | 0.57 | 0.56 |
| **VGG16** | *MGMT-Methylated* | 0.60 | 0.70 | 0.65 |
|  | *MGMT-Unmethylated* | 0.60 | 0.48 | 0.53 |

**Supplementary Table S2.** Detailed performance metrics of two state-of-the-art (SOTA) deep learning (DL) architectures, ConvNeXtTiny and EfficientNetV2S, for predicting MGMT promoter methylation status in GBM tumors. The table reports class-specific precision, recall, and F1-score values for each model.

|  |  | **Precision** | **Recall** | **F1-Score** |
|---|---|---|---|---|
| **ConvNeXtTiny** | *MGMT-Methylated* | 0.52 | 1.0 | 0.69 |
|  | *MGMT-Unmethylated* | 0.1 | 0.1 | 0.1 |
| **EfficientNetV2S** | *MGMT-Methylated* | 0.54 | 0.89 | 0.67 |
|  | *MGMT-Unmethylated* | 0.59 | 0.18 | 0.27 |